\documentclass[prx,twocolumn,showpacs]{revtex4}
\usepackage{graphicx}
\usepackage{amsmath}

\usepackage{amsmath}
\usepackage{amsfonts}

\usepackage{epsfig} 

\usepackage{epstopdf}
\newcommand{\Ket}[1]{\left|#1  \right>}

\newcommand{\Braket}[1]{\left<#1  \right>}

\def\begeq{\begin{equation}}
\def\endeq{\end{equation}}

\def\begeqar{\begin{eqnarray}}
\def\endeqar{\end{eqnarray}}

\usepackage{color}

\begin{document}

\title{Universal nonequilibrium signatures of Majorana zero modes in quench dynamics}

\author{R. Vasseur$^{1,2}$, J. P. Dahlhaus$^{1}$, and J. E. Moore$^{1,2}$}
\affiliation{$^1$Department of Physics, University of California, 
Berkeley, California 95720, USA}
\affiliation{$^2$Materials Science Division, Lawrence Berkeley National Laboratory, Berkeley CA 94720, USA}

\date{\today}

\begin{abstract}

The quantum evolution after a metallic lead is suddenly connected to an electron system contains information about the excitation spectrum of the combined system.  We exploit this type of ``quantum quench'' to probe the presence of Majorana fermions at the ends of a topological superconducting wire.  We obtain an algebraically decaying overlap (Loschmidt echo) ${\cal L}(t)=\left| \Braket{\psi(0) | \psi(t)} \right|^2\sim t^{-\alpha}$  for large times after the quench, with a universal critical exponent $\alpha=\frac{1}{4}$ that is found to be remarkably robust against details of the setup, such as interactions in the normal lead, the existence of additional lead channels or the presence of bound levels between the lead and the superconductor. As in recent quantum dot experiments, this exponent could be measured by optical absorption, offering a new signature of Majorana zero modes that is distinct from interferometry and tunneling spectroscopy.
\end{abstract}


\maketitle

\section{Introduction}

The original example of Anderson's orthogonality catastrophe~\cite{andersonorthogonality} was a vanishing overlap between two quantum states of a large number of non-interacting electrons, one with and one without a localized impurity potential $V(r)$. This phenomenon governs the electronic response when X-rays are absorbed in a metal~\cite{XRay1,XRay2}: the core hole generated in the absorption process acts as a localized potential, and the time-dependent response of the system after this change in the Hamiltonian was perhaps the first nontrivial example of a quantum quench in a many-electron system.  Quantum quenches have been of great interest recently as a basic question about non-equilibrium physics appearing in many contexts~\cite{Calabrese:2006,Lamacraft:2007p841,Lamacraft2012177,Latta,Caux,Rigol}.

The point of the present work is to study a quantum quench in an electron system that supports Majorana fermion excitations~\cite{RevZhang,RevAlicea,RevBeenakker}, specifically a topological superconducting nanowire~\cite{Kitaev,Oreg,Lutchyn} of the type sought in recent experiments~\cite{Delft,Deng,Rokh, Das, Marcus, Finck,Franceschi}.  The quench consists of suddenly connecting the nanowire to an ordinary metallic lead.  The behavior of the many-electron wavefunction at long times after the quench is significantly altered by the presence of the Majorana excitation: the wavefunction overlap with the initial state (the Loschmidt echo) decays with a universal power-law, unlike in the Anderson orthogonality case where the exponent is non-universal.  This effect of the Majorana fermion can be distinguished from effects of ordinary fermions, either trapped or extended, and can be understood as resulting from an induced change in the effective boundary condition of the ordinary fermions in the metallic lead.

The detailed analysis of the long-time behavior after the quench is possible because the quench's effect of changing the boundary condition on the metallic lead, from normal reflection to Andreev reflection,  is represented by a known operator in boundary conformal field theory (BCFT)~\cite{Cardy}.  We also find that interactions in the metallic lead, which produce a Luttinger-liquid state, do not modify the universal exponent indicating a Majorana fermion -- in the same regime of parameters where the zero-bias conductance anomaly is stable, and similarly the result is impervious to the presence of additional channels or localized states.  We confirm the predictions of the field theory numerically, both by free fermion methods and density-matrix renormalization group (DMRG) simulations~\cite{White, Vidal1, Vidal2,Schollwoeck}.

While much of our presentation focuses on the basic phenomena resulting from the quantum quench, recently a similar quench of tunneling into a quantum dot in the Kondo regime was achieved experimentally by optical absorption~\cite{Latta}.  We discuss some conditions for a possible experiment using optical absorption in a (non-Kondo) quantum dot.  We believe that such a measurement is conceptually distinct from previously proposed detection methods for solid-state Majorana excitations including e.g. interferometry, tunneling spectroscopy, current noise, and the $4\pi$ periodic Josephson effect~\cite{Sengupta,Bolech,Nilsson,LawLee,Fu2,FuKane,Wimmer,Akhmerov}. The presence of Majorana fermions in a system results in a strong modification of the absorption edge singularity that is one of the classic features of metallic electrons. The universal nature of the Majorana signature can serve to distinguish it from other processes like e.g. the Kondo effect or consequences of disorder, which depend on the experimental parameters of the setup.

The remainder of this paper is organized as follows.  Section II introduces the basic model of Kitaev~\cite{Kitaev} of a spinless $p$-wave superconducting wire that in one phase supports Majorana edge zero modes.  At time $t=0$, the wire is suddenly tunnel-coupled to the end of a non-interacting metallic 1D wire in its ground state.  The long-time dynamics is analyzed using the low-energy limit of the coupled wires, and the decay of the Loschmidt echo is determined in a boundary conformal field theory approach by the scaling dimension of the operator that changes boundary conditions on the normal wire.  The change of boundary conditions can be derived from observing that, when the normal wire is written in terms of a pair of Majorana fermions, one of the Majorana fermions undergoes a phase shift as a result of Andreev reflection at the junction.  The resulting predictions are numerically confirmed for both ordinary and topological phases of the model.

Section III shows that the Majorana effect on the long-time overlap exponent is as stable to interactions in the lead as the zero-bias tunneling conductance anomaly, and also to the presence of additional channels, which is an important factor in current experiments. Some of these predictions are also verified numerically using DMRG simulations. Section IV describes some features of a possible experiment using the optical absorption of a quantum dot between the normal lead and the superconducting wire, and discusses effects of additional localized electron states as modeled by such a quantum dot. Finally, section V provides a discussion of the results and explains how the theoretical analysis of the quench in terms of a boundary condition change can be generalized to other kinds of topological 1D systems, including those with parafermionic excitations.

\section{Majorana-induced quench dynamics in a non-interacting metallic lead}
\label{secKitaev}

\subsection{Spinless normal metal -- superconductor junction}

We consider the quench dynamics of a normal lead (NL) suddenly coupled to a (topological) superconductor (TSC). To illustrate our main ideas, we start our discussion with a simple spinless $p$-wave superconductor as introduced by Kitaev~\cite{Kitaev}
\begin{align}
H_{\rm SC} &= - J \sum_{i=1}^{L-1} \left( f^\dagger_{i+1} f_i +{\rm h.c.}\right) - \mu \sum_{i=1}^{L} \left( f^\dagger_{i} f_i -\frac{1}{2}\right) \notag\\
&+ \Delta_s \sum_{i=1}^{L-1}  \left( f^\dagger_{i+1} f^\dagger_i + f_i f_{i+1}  \right),
\label{eqLattice1}
\end{align}
with uniform positive coupling $J$ and real superconducting gap $\Delta_s$ (superconducting phase $\phi=0$). This so-called Kitaev chain is tunnel-coupled ($J'\ll J$)
\begin{equation}
H_{\rm t} = -J^\prime \left( c^\dagger_1 f_1 + {\rm h.c.}\right),
\label{eqLattice2}
\end{equation}
to a non-interacting metallic lead
\begin{equation}
H_{\rm L} = - J\sum_{i=1}^{L-1} \left( c^\dagger_{i+1} c_i +{\rm h.c.}  \right).
\label{eqNormalLead}
\end{equation}
The dispersion relation of the Kitaev chain reads $\omega_k = \pm \sqrt{(2 J\cos k +\mu)^2 + 4 \Delta_s^2 \sin^2 k}$ and the problem is known to have a quantum phase transition at $\left|\mu \right|= 2 J$, separating a topologically trivial phase $(\left|\mu \right| > 2 J$) from a topological phase that hosts Majorana zero modes at the edges ($\left|\mu \right| < 2 J$). 

\begin{figure}
\hspace{-1.2cm}
    \includegraphics[scale=.24]
    {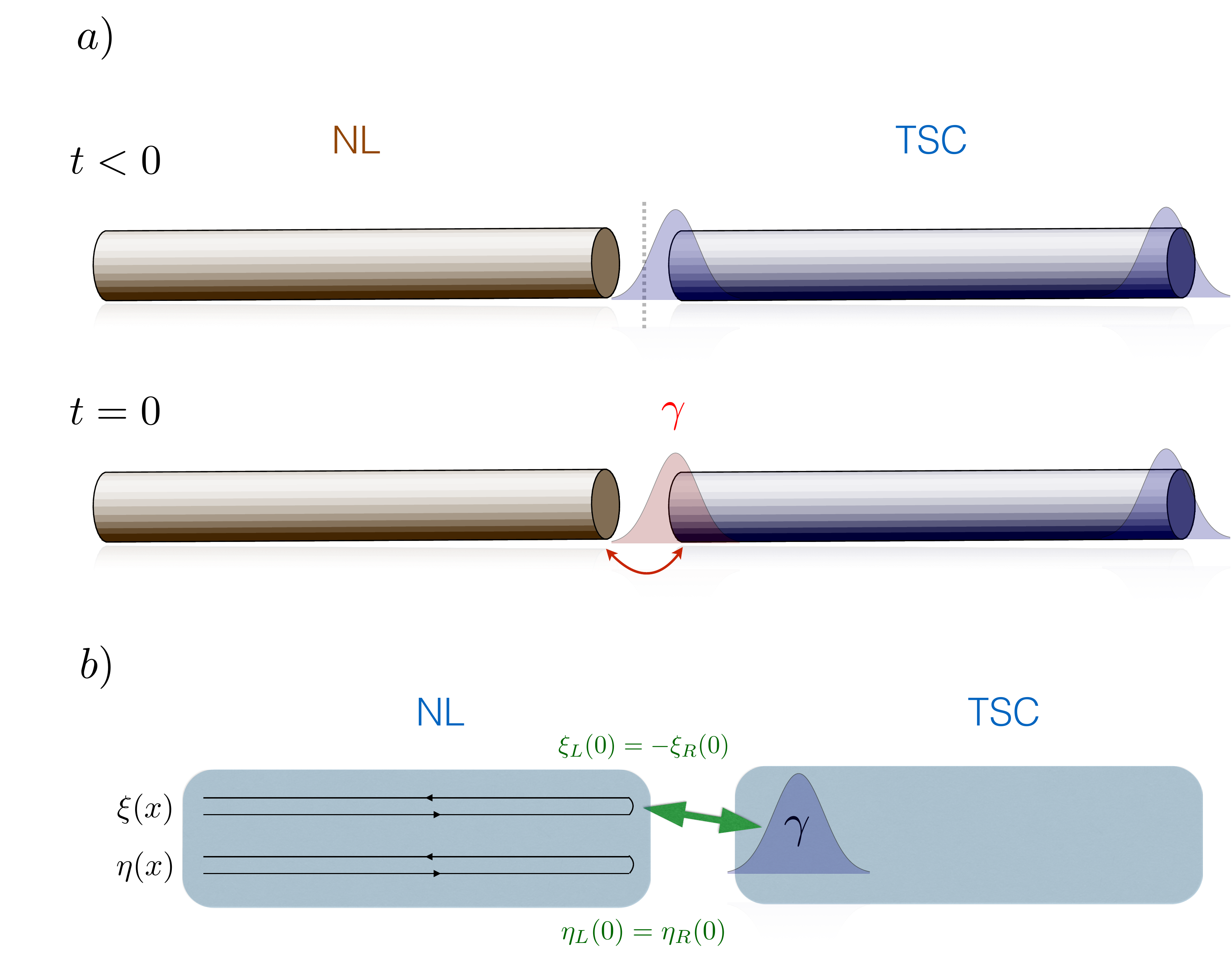}
    \caption{Physical setup. a) Local quantum quench considered in this paper: at time $t=0$ a one-dimensional metallic (normal) lead is connected to the end of a topological superconductor with Majorana zero modes at both ends. b) Cartoon representation of the low energy fixed point (large time behavior) for a single-channel non-interacting lead. Only one of the Majorana channels in the normal lead $\psi(x)=(\xi(x)+i \eta(x))/2$ experiences a $\pi/2$ phase shift corresponding to the Andreev reflection $\psi^\dagger_R(0)=\psi_L(0)$ at the junction. }
\label{FigSetup}
\end{figure}

We initially prepare the system in the ground state $\Ket{\psi_0} = \Ket{\Omega}_{\rm SC} \otimes \Ket{\Omega}_{\rm L}$ for decoupled lead and superconductor ($J'=0$). At time $t=0$, the coupling $J'$ is suddenly turned on so that the system is brought far from equilibrium (Fig.~\ref{FigSetup}-a). The wave function at time $t>0$ is simply given by the unitary evolution $\Ket{\psi(t)}= {\rm e}^{-i H t}\Ket{\psi_0}$ with $H=H_{\rm SC} +H_{\rm L} +H_{\rm t}   $.

Most of our discussion will focus on the behavior at large times, which is dominated by the only low energy excitation of the topological superconductor -- the Majorana zero mode.  In what follows, we require that the time be larger than the inverse of the band-width such that a field theoretic continuum theory applies. Assuming that the massive degrees of freedom in the superconductor can be integrated out then yields an effective Majorana boundary term for the metallic lead. 

This coupling to a Majorana fermion introduces a typical energy scale that shall be denoted by $T^\star$ in the following. In a finite system, the overlap of the Majorana modes from opposite ends of the Kitaev chain leads to another small (yet non-zero) energy scale $\delta_M$ for the Majorana excitations. We expect to start seing the effects of this overlap in the post-quench dynamics for large times $t>t_1\sim1/\delta_M$. In the following we will assume the thermodynamic limit $L\rightarrow \infty$ though, where the time scale $t_1\rightarrow \infty$ diverges due to the exponentially decaying overlap of the Majorana modes ($\delta_M\sim {\rm e}^{ -\alpha L}$). Thus we can safely set $\delta_M=0$ and be left with a single energy scale $T^\star$ in the problem -- the normal lead being scale invariant in the scaling limit. 

\subsection{Low energy description and flow from free to fixed boundary conditions in the Ising model}

Given these prerequisites, we can linearize the dispersion relation of the metallic lead near the Fermi energy. The Hamiltonian~\eqref{eqNormalLead} then takes the familiar form 
\begin{equation}
H_{\rm L}=  - i v_F \int_{0}^\infty {\rm d} x\; \left( \psi_R^\dagger \partial_x \psi_R - \psi_L^\dagger \partial_x \psi_L \right),
\end{equation}
with the Fermi velocity $v_F=2 J$ at half-filing. In order to analyze the effective boundary terms due to the superconductor, it is very convenient to  ``unfold'' the normal lead and define a right moving field $\psi(x)$ on the real line: $\psi(x)=\psi_R(x)$ for $x>0$ and $\psi(x)=\psi_L(-x)$ for $x<0$. When the Kitaev chain is in the topological phase ($\left| \mu \right| < 2 J$), and assuming that it is fully gapped throughout, we can integrate out the massive degrees of freedom of the superconductor to obtain the low-energy Hamiltonian for the chiral fermonic field $\psi(x)$
\begin{equation}
H = - i v_F \int_{-\infty}^\infty {\rm d} x\; \psi^\dagger \partial_x \psi + i \kappa \gamma \left( \psi^\dagger+\psi  \right)(0),
\label{FFmajorana}
\end{equation}
which describes a Majorana zero mode $\gamma$ ($\gamma^2=1$ and $\gamma^\dagger =\gamma$), coupled with strength $\kappa\propto J^\prime$ to the fermion $\psi(x)$ describing the normal lead, with regularization $\psi(0)\equiv\frac{\psi(0^-)+\psi(0^+)}{2}$. Note that in principle eq.~\eqref{FFmajorana} contains other effective boundary terms (less relevant in the renormalization group sense) that turn out to be important only in the absence of the Majorana zero mode --- {\it i.e.} when the superconductor is in the trivial phase (see Section~\ref{SecTopTrivial} below). For the sake of simplicity, we will ignore these other terms in the following.

To proceed, we then decompose $\psi$ into Fourier modes
\begin{equation}
\psi(x,t) = \int {\rm d} \omega \ {\rm e}^{i \omega (x-  t)} \psi_{\omega} (x),
\end{equation}
where we set $v_F=1$. The diagonalization of this non-interacting problem can be expressed in terms of a scattering matrix
\begin{equation}
\left( \begin{array}{c} 
\psi_{\omega}(0^+) \\
\psi^\dagger_{-\omega}(0^+)
\end{array} \right) = \hat{S}(\omega) \left( \begin{array}{c} 
\psi_{\omega}(0^-) \\
\psi^\dagger_{-\omega}(0^-)
\end{array} \right).
\label{eqDefSMatrix}
\end{equation}
The precise form of the matrix $ \hat{S}(\omega)$ is not important (see e.g. Ref.~\cite{MajoranaLuttinger} for a related calculation). What matters is that since $T^{\star}=\kappa^2$ is the only energy scale of the problem, $ \hat{S}$ has to be a function of $\omega/T^{\star}$. In the Renormalization Group (RG) language, this amounts to saying that the boundary perturbation in eq.~\eqref{FFmajorana} has dimension $\Delta=\frac{1}{2}$ and is therefore relevant.  From the explicit form of $ \hat{S}(\omega)$ , one finds that at high energy ($\omega \gg T^{\star}$),  $ \hat{S}(\omega)$ is the identity matrix: this corresponds to the boundary condition $\psi(0^+)=\psi(0^-)$ for the fermion in the lead (or $\psi_R(0)=\psi_L(0)$ before folding).  At low energy ($\omega \ll T^{\star}$), the relevant boundary perturbation $ i \kappa \gamma \left( \psi^\dagger+\psi  \right)(0)$ drives the system into a new RG fixed point (or more precisely, a new conformally invariant boundary condition)
\begin{equation}
\hat{S}(0) = \left( \begin{array}{cc} 
0 & -1\\
-1 & 0
\end{array} \right),
\end{equation}
which corresponds to the Andreev reflection condition $\psi^\dagger(0^+)=\psi(0^-)$ after the canonical transformation $\psi\rightarrow i\psi$, $\psi^\dagger \rightarrow - i\psi^\dagger$. This boundary condition governs the low energy properties of the junction, and in particular, yields the zero-bias tunneling conductance $G=2 e^2/h$~\cite{LawLee}. 

In our context of quantum quenches, it is very instructive to analyze this non-interacting setup purely in the language of Majorana fermions. Let us introduce $\psi=\frac{\xi + i \eta}{2}$, with $\xi$ and $\eta$ real Majorana fermions:  $\lbrace \xi (x), \xi (x^\prime) \rbrace= 2\delta(x -x^\prime)$ and $\lbrace \eta (x), \eta (x^\prime) \rbrace= 2\delta(x -x^\prime)$. In terms of these Majorana fermions, eq.~\eqref{FFmajorana} now reads
\begin{align}
H &= - \frac{i}{4} \int_{0}^\infty {\rm d} x \left( \xi_R \partial_x \xi_R - \xi_L \partial_x \xi_L \right) \notag \\
&- \frac{i}{4} \int_{0}^\infty {\rm d} x \left( \eta_R \partial_x \eta_R - \eta_L \partial_x \eta_L \right) + i \kappa \gamma  \xi(0).
\end{align}
Our scattering problem can therefore be mapped onto two Majorana chains (or equivalently two independent Ising models), where one of the copies, $\eta(x)$, decouples from the boundary Majorana $\gamma$ and thus does not contribute to the dynamics. The remaining Hamiltonian can be identified with an Ising spin chain with a boundary magnetic field $\kappa$ which induces a flow from free to fixed boundary conditions. The equations of motion can readily be solved in frequency space in terms of the scattering matrix $\xi_\omega(0^+)=S_\xi(\omega)\xi_\omega(0^-)$ with
\begin{equation}
\label{eqScattKondo}
S_\xi(\omega) = \frac{i \omega - T^\star/2}{i \omega + T^\star/2},
\end{equation}
so that  $\xi(0^+)=-\xi(0^-)$ and $\eta(0^+)=\eta(0^-)$ at low energy, corresponding to full Andreev reflection (Fig.~\ref{FigSetup}-b). The boundary entropy drop~\cite{gFactor} associated with this RG flow is simply given by $\ln \sqrt{2}$, where $d=g_{\rm UV}/g_{\rm IR}=\sqrt{2}$ is the quantum dimension of the Majorana fermion $\gamma$ that was free at high energy and that becomes fully  hybridized with the normal lead at low energy. This will turn out to have crucial consequences on the quench dynamics of the system.

\subsection{Quench dynamics and Loschmidt echo}

In this paper, we will characterize the dynamics of the system using the time-dependent overlap (fidelity, or Loschmidt echo) ${\cal L}(t)=\left| \Braket{\psi(0) | \psi(t)} \right|^2$ that encodes how far the system is from its initial state at a given time $t$. Based on pure scaling, it is natural to expect  ${\cal L}(t)$ to be a universal function of $t T^{\star}$ only~\cite{QuenchBCC}, since $t$ acts effectively as the inverse of a typical energy scale. For large times, the fact that the (boundary of the) system is flowing to a completely new (boundary) RG fixed point leads to an algebraic decay ${\cal L}(t)  \sim t^{-\alpha}$ that can be interpreted as a time dependent version of the celebrated Anderson orthogonality catastrophe~\cite{andersonorthogonality}. The corresponding critical exponent can be conveniently computed using Boundary Conformal Field Theory (BCFT). The key idea is to interpret the  Loschmidt echo in imaginary time
\begin{equation}
{\cal L}(t= -i \tau) = \left |\Braket{\psi_0 | {\rm e}^{- H \tau} | \psi_0  } \right|^2,
\end{equation}
as the square of a partition function of a 2D statistical problem at its critical point -- two copies of the Ising model in our case -- in the half-plane $x \in [0,  \infty)$, $y \in (-\infty,  \infty)$. The quantum quench then amounts to changing the boundary condition at $x=0$ -- {\it i.e.} applying some sort of boundary magnetic field -- for  $y \in [0,  \tau]$. For large $ \tau$, this boundary condition flows to a conformally invariant one and this geometry can be thought of as the two-point function of a Boundary Condition Changing (BCC) operator in the BCFT language~\cite{Cardy}. This leads to the identification $\alpha=4 h_{\rm BCC}$~\cite{QuenchBCC}, where $h_{\rm BCC}$ is the dimension of the corresponding BCC operator (see Appendix~\ref{AppLoschBCC} for more details). This formula $\alpha=4 h_{\rm BCC}$ is completely general and allows us to rely on well-established conformal field theory results to compute the asymptotic behavior of ${\cal L} (t)$. Note however that marginal perturbations have to be treated separately as they modify $\alpha$ in a continuous fashion (see sections~\ref{SecTopTrivial} and~\ref{Secmultichannel} below).

Applied to our problem, this line of reasoning yields
\begin{equation}
{\cal L}(t) \underset{t \gg ( T^{\star})^{-1}}{\sim} t^{-1/4},
\label{eqExpG}
\end{equation}
where the exponent $1/4 = 4 h_{\rm BCC}$ corresponds to an operator changing boundary conditions from $\psi(0^+)=\psi(0^-)$  to  $\psi^\dagger(0^+)=\psi(0^-)$ in the $c=1$ Dirac fermion theory, with dimension $h_{\rm BCC} = \frac{1}{16}$. Using the Ising formulation of the previous section, the exponent $h_{\rm BCC} = \frac{1}{16}$ can also be thought of as the dimension of the spin operator, which is well-known to correspond to changing boundary conditions from free to fixed in the Ising model~\cite{Cardy}. 

Note that the exponent resulting from this boundary condition is half as large as one would obtain from the boundary condition of a $\delta=\pi/2$ phase shift on a normal fermion, as here only one of the two Majorana degrees of freedom picks up that phase shift $\mathrm{e}^{2i \delta}=-1$.  If charge were conserved (which is not the case in the present model because of the superconductor), this exponent would correspond to flow of one-half an electron charge to the vicinity of the boundary~\cite{chargeflow1,chargeflow2}.

The power-law dependence~\eqref{eqExpG} is characteristic of Majorana zero modes, and we shall argue in the remainder of this paper that it holds for more realistic systems, including several channels and interactions in the metallic lead.

\begin{figure}
\centering
    \includegraphics[scale=.44]
    {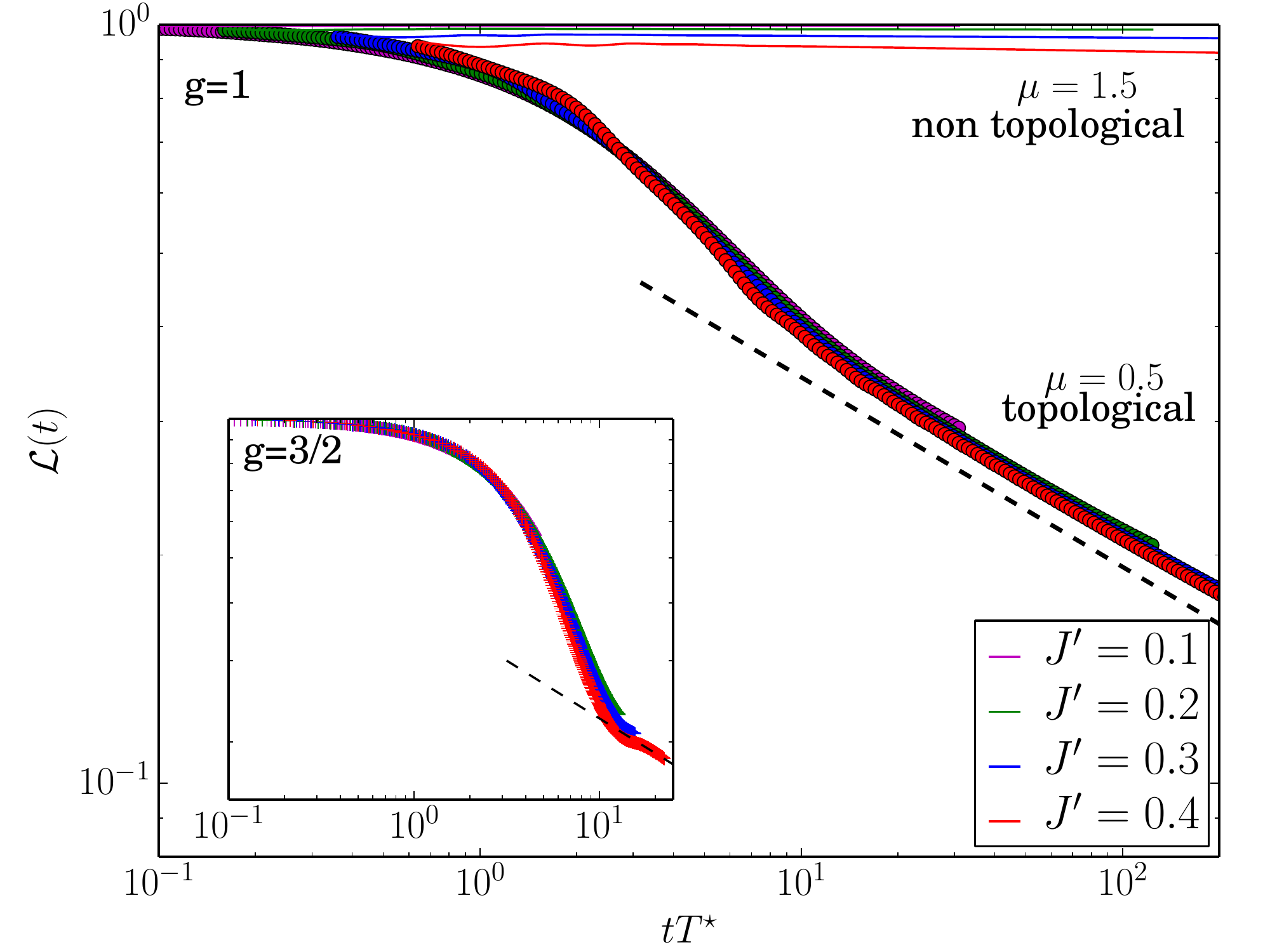}
    \caption{Loschmidt echo for a quench of the tunneling between a non-interacting normal lead and the Kitaev model for a spinless 1D p-wave superconductor. We work with $J=\frac{1}{2}$, $\Delta_s=1.0$ and $L=4000$. When the superconductor is topologically non-trivial, the Majorana zero modes at its edges induce a universal decay $t^{-1/4}$ of the Loschmidt echo (dashed line). The data for different values of the tunneling $J^\prime$ collapse once properly rescaled by $T^\star = (J^\prime/J)^{(2g/(2g-1))}$. Inset: Loschmidt echo for an interacting normal lead with Luttinger parameter $g=\frac{3}{2}$, $\mu=0.5$, $\Delta_s=2.5$ and $L=200$   (total system size $N=2L=400$ sites) from DMRG.}
\label{FigLosch1}
\end{figure}

\subsection{Topologically trivial case}

\label{SecTopTrivial}

When the superconductor is in a topologically trivial phase, it is natural to expect the Loschmidt echo to remain unity in the limit of infinite gap -- in other words, nothing happens when the tunneling is suddenly turned on. In actual systems however (and in the numerical simulations that will be described bellow), the gap in the superconductor is finite. It is thus crucial to understand the dynamics for a metallic lead suddenly coupled to a gapped phase with gap $\Delta$, be it a superconductor or not. Integrating out the gapped phase then yields the effective boundary term for the metallic lead
\begin{equation}
H = - i v_F \int_{-\infty}^\infty {\rm d} x \psi^\dagger \partial_x \psi + V(\Delta)  \psi^\dagger(0) \psi(0),
\label{eqNonTopo}
\end{equation}
where the amplitude of the boundary perturbation can be estimated in perturbation theory as $V(\Delta) \sim (J^\prime)^2 / \Delta$.  It turns out that superconductivity is not important in the context of this single-channel model, as Cooper pair tunneling processes, described by an effective boundary term $i \delta(x) \psi \partial \psi  + {\rm h.c.}$~\cite{MajoranaLuttinger}, are irrelevant at low energy (large times). This term and other ones involving higher order derivatives were thus dropped in~\eqref{eqNonTopo}. 

The potential scattering term in~\eqref{eqNonTopo} can be thought of as a singular gauge potential $A(x)=V(\Delta) \delta(x)$ which can be gauged away by enforcing a phase shift $\psi(0^+) = \mathrm{e}^{i V(\Delta)}\psi(0^-)$. This phase shift yields a power-law decay of the Loschmidt echo ${\cal L}(t) \sim t^{-\alpha}$, with an Anderson orthogonality exponent $\alpha=\left( \frac{V(\Delta)}{2\pi}\right)^2$ that can readily be computed from bosonization for instance. Therefore, for a quench involving a non-topological superconductor with large yet finite gap $\Delta$ (or any ordinary gapped phase for that matter), we expect the large time scaling
\begin{equation}
{\cal L}(t) \sim t^{-{\rm Const}/\Delta^2},
\label{eqExpGNonTopo}
\end{equation}
where the (small) exponent is non-universal. Importantly, in the presence of Majorana zero modes, boundary terms such as $V(\Delta)  \psi^\dagger(0) \psi(0)$ do not influence the universal result~\eqref{eqExpG}.

\begin{figure}
\centering
    \includegraphics[scale=.43]
    {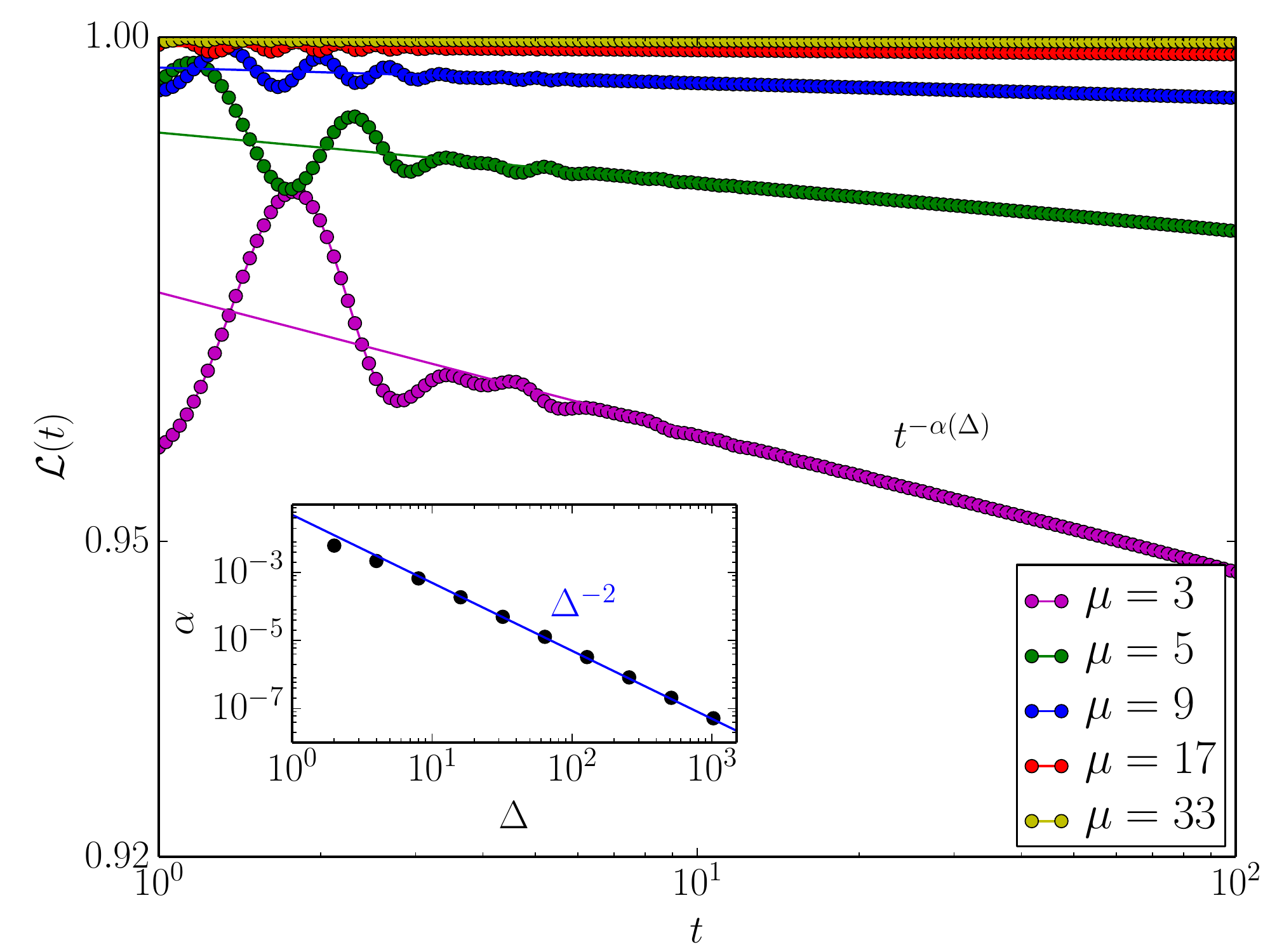}
    \caption{Loschmidt echo after a quench of the tunneling between a (non-interacting) normal lead and a topologically trivial gapped phase with gap $\Delta$. We choose $L=4000$, $J'=\frac{1}{2}$, and $\Delta_s=0$ to obtain a trivial band insulator with gap $\Delta=\mu-1$. The Loschmidt echo exhibits a clear power-law behavior as a function of time with a small exponent that scales as $\Delta^{-2}$ (Inset). Results for a non-zero superconducting gap $\Delta_s \neq 0$ show exactly the same physics.}
\label{FigLosch2}
\end{figure}

\subsection{Numerics}

In order to check our predictions for the large time dynamics, we compute numerically the Loschmidt echo for the Hamiltonian $H=H_{\rm L}+H_{\rm SC}+H_{\rm t}$ given by eqs~\eqref{eqLattice1},~\eqref{eqLattice2},~\eqref{eqNormalLead}. Since the system is non-interacting, the Loschmidt echo can be expressed as a determinant thus allowing a computation on fairly large systems. We refer the reader to Ref.~\cite{Ste11} and to appendix~\ref{AppNumerics} for technical details on the numerical method.

 Results of the simulations are shown in Fig.~\ref{FigLosch1} (numerical parameters are given in the caption). The thick lines/dots are for $\mu=0.5$ and different values of $J^\prime$, where the superconductor is in the topological phase and supports Majorana fermions at its ends. The double logarithmic plot of Loschmidt echo versus rescaled time $t T^\star = t (J'/J)^2$ demonstrates the predicted universal power law decay $\mathcal{L}(t)\sim t^{-1/4}$ (compare dashed line), with nicely collapsing curves for different values of $J'$. Both the universal collapse and the power-law behavior are signatures of the presence of Majorana zero-modes at the ends of the Kitaev chain in this regime.

In the case of a trivial superconductor, $\mu=1.5$ (thin lines), the curves do not collapse and show very slow decays that can be explained by the fact that the gap in the superconductor is large but finite. Fig. \ref{FigLosch2} shows that these decays follow power laws $\sim t^{-\alpha(\Delta)}$, with small exponents $\alpha(\Delta)$ depending on the actual gap $\Delta$ of the system. For large $\Delta$, this dependence is $\alpha\sim\Delta^{-2}$ as demonstrated in the inset, consistent with the prediction~\eqref{eqExpGNonTopo}.

\section{Stability of the exponent}

\subsection{Effect of interactions in the metallic lead}

A remarkable feature of~\eqref{eqExpG} is that it is robust, with the same exponent, against quite strong interactions in the lead. To demonstrate this, we add an interaction between electrons on adjacent sites
\begin{equation}
H_{\rm I} = U\sum_{i=1}^{L-1}  \ \left( c^\dagger_{i}c_i -\tfrac{1}{2}\right)  \left( c^\dagger_{i+1}c_{i+1} -\tfrac{1}{2}\right), 
\label{eqNormalLeadInt}
\end{equation}
 to the lead Hamiltonian in Eq. (\ref{eqNormalLead}). In the scaling limit, the lead can then be described in terms of a Luttinger liquid  with Luttinger parameter $g^{-1}= 2-\frac{2}{\pi} \arccos U$~\cite{LuttingerParameter}.  In the corresponding low energy description, the unfolded chiral (right-moving) fermionic field $\psi(x)$ can be bosonized~\cite{Giamarchi} as $\psi(x)=\frac{\chi}{\sqrt{2 \pi}} \mathrm{e}^{i  \phi}$ where $\chi$ is a Klein factor -- yet another Majorana fermion -- introduced to make sure that $\psi(x)$ anti-commutes with $\gamma$ (see also~\cite{MajoranaLuttinger} for a discussion of the bosonization of the Majorana boundary interaction). The Hamiltonian of the interacting lead then reads
\begin{equation}
H_{\rm L} = \frac{g}{4 \pi} \int {\rm d} x (\partial_x \phi)^2 , 
\label{eqNormalLeadBos}
\end{equation}
while the coupling to the Majorana becomes
\begin{equation}
i \kappa \gamma \left( \psi^\dagger+\psi  \right)(0) \sim \kappa \sigma_x \cos  \phi,
\label{eqBsG}
\end{equation}
after bosonization, where we have introduced the Pauli matrix representation $\sigma_x=i \gamma \chi$. This perturbation has dimension $\Delta=\frac{1}{2g}$ and is hence relevant if $g>\frac{1}{2}$~\cite{MajoranaLuttinger,AffleckGiuliano} (see also~\cite{LutchynSkrabacz} for an example of application). For $g<\frac{1}{2}$, the Majorana term is irrelevant and the universal behavior of the Loschmidt echo is lost, just as the quantized zero-bias conductance.

When the perturbation is relevant, the corresponding energy scale induced by the perturbation is given by $T^{\star} \sim \kappa^{2g/(2g-1)}$ and one still expects the dynamics to scale as $t T^{\star}$. Like in the free fermion case, the effective boundary condition for the lead excitations flows from $\psi(0^+)=\psi(0^-)$  to  $\psi^\dagger(0^+)=\psi(0^-)$, which correspond respectively to Neumann and Dirichlet boundary conditions for the boson $\phi$. This is consistent with the boundary sine-Gordon interaction~\eqref{eqBsG} which pins down the value of $\phi(0)$ as $\kappa \to \infty$. The associated boundary condition changing operator has dimension $h_{\rm BCC} = \frac{1}{16}$ regardless of the Luttinger parameter~\cite{BCC_ND1,BCC_ND2}, so that Eq.~\eqref{eqExpG} holds for an interacting lead as well.

\subsection{t-DMRG results }

In order to check that signatures of Majorana persist in the presence of interactions in the lead (with $g>1/2$), we use a Density Matrix Renormalization Group (DMRG) algorithm~\cite{White} to simulate the time evolution~\cite{Vidal1,Vidal2} in terms of matrix product states (MPS)~\cite{Schollwoeck}. We adapt the bond dimension $\chi$ of the MPS in order to keep the discarded weight $\epsilon$ below $10^{-7}$ throughout the whole time evolution with a Trotter time step $dt=0.1$ and a fourth-order Trotter decomposition. We stop the simulations when $\chi \sim {\cal O}(700)$. Results for $U=-0.5$, corresponding to a Luttinger parameter $g=\frac{3}{2}$, are shown in the inset of Fig.~\ref{FigLosch1}. Even though the rapid build up of entanglement in the system makes it difficult to access the power-law regime~\eqref{eqExpG}, we observe a clear collapse of our numerical data for different values of $J^\prime$ when rescaled using $T^{\star} \sim (J^\prime)^{2g/(2g-1)}$ consistent with our expectations. Repulsive interactions ($U>0$, $g<1$) unfortunately seem to require to work with larger superconducting gaps and smaller time steps in order to converge and observe the same physics, thus making this real time DMRG approach rather impractical. We expect that recently introduced DMRG approaches aimed at computing the Fourier transform of the Loschmidt echo $\Braket{\psi(0)|\psi(t)}$ rather than the echo itself (see {\it e.g.}~\cite{MajoDMRG} for a related calculation) may be more efficient to extract the large time behavior~\eqref{eqExpG}.

\subsection{Interacting case with two channels (spinful)}

In an actual experiment, the wire contains several channels, and it is natural to wonder if our prediction also applies to that case. We first set out to investigate the effect of a second lead channel in the presence of interactions in the lead. In the next section, we turn to the full multichannel case, restraining ourselves from a discussion of interaction effects though.

Let us consider an interacting two-channel lead (spinful Luttinger liquid) where we think of the two channels as corresponding to two spin states $\sigma=\uparrow, \downarrow$. Note that in the presence of spin orbit coupling, the labeling of the channels as being spin up and down could be slightly misleading, but we will nevertheless refer to the two channels in this way for convenience. It can be described by the Luttinger liquid Hamiltonian
\begin{equation}
H_{\rm L} = \int_0^\infty {\rm d}x \sum_{\alpha=c,s} \frac{v_{\alpha}}{2} \left(\frac{1}{K_\alpha} (\partial_x \Phi_\alpha)^2 + K_\alpha (\partial_x \theta_\alpha)^2  \right),
\label{eq:spinfullead}
\end{equation}
where $\alpha=c,s$ labels the charge and spin modes of the fields $\Phi_\alpha$ and $\theta_\alpha$, with velocity $v_\alpha$ (set to unity in the following), and Luttinger parameter $K_\alpha$. 

We will concentrate here on the case where the lead has SU$(2)$ symmetry ($K_{s}=1$) -- ignoring spin orbit coupling for the sake of simpler arguments. For the generic case of a two-channel Luttinger liquid with broken SU$(2)$ symmetry similar conclusions can be drawn. 

The coupling to a topological superconductor has been studied in details in~\cite{MajoranaLuttinger,AffleckGiuliano} and was argued to drive the system to a RG fixed point dubbed $A \otimes N$ corresponding to Andreev reflection for, say, the up channel, while the down channel experiences normal reflection. In the bosonization language, this corresponds to the conformally invariant boundary condition $\Phi_{\uparrow}(0)=0$, $\theta_{\downarrow}(0)=0$. In terms of the spin and charge modes, this yields $\Phi_c(0)=-\Phi_s(0)$ and $\theta_s(0)=\theta_c(0)$. In order to understand the large time dynamics of the system after the quantum quench, we need to diagonalize simultaneously the bulk and the boundary conditions. The bulk can be easily diagonalized by introducing the new right and left movers $\Phi_c = \sqrt{K_c} (\varphi^R_c+\varphi^L_c)$,  $\Phi_s = \varphi^R_s+\varphi^L_s$, $\theta_c = \frac{1}{\sqrt{K_c}} (\varphi^R_c-\varphi^L_c)$,  $\theta_s = \varphi^R_s-\varphi^L_s$ so that the Hamiltonian reads $H_{\rm L} = \int_0^\infty {\rm d}x \sum_{\alpha=c,s} \left[ (\partial_x \varphi^R_\alpha)^2+(\partial_x \varphi^L_\alpha)^2  \right]$. The low energy $A \otimes N$ boundary condition at $x=0$ becomes however fairly complicated 
\begin{equation}
\left( \begin{array}{c} 
\varphi^L_c (0)\\ \varphi^L_s(0) 
\end{array} \right)
 = \left( \begin{array}{cc} 
\frac{1-K_c}{1+K_c} & -\frac{2 \sqrt{K_c}}{1+K_c} \\
-\frac{2 \sqrt{K_c}}{1+K_c} & \frac{K_c-1}{1+K_c} 
\end{array} \right) \left( \begin{array}{c} 
\varphi^R_c (0)\\ \varphi^R_s(0) 
\end{array} \right).
\end{equation}
This boundary condition can be diagonalized by a unitary transformation that leaves the bulk Hamiltonian unchanged
\begin{align}
\phi_1^{R/L} &= \frac{1}{\sqrt{1+K_c}} \left(\sqrt{K_c} \varphi^{R/L}_c + \varphi^{R/L}_s \right), \notag \\
\phi_2^{R/L} &= \frac{1}{\sqrt{1+K_c}} \left(-\varphi^{R/L}_c + \sqrt{K_c} \varphi^{R/L}_s\right).
\end{align}
It is easy to check that these new bosons satisfy $A \otimes N$ boundary conditions as well: $\phi^R_1(0)=-\phi_1^L(0)$, $\phi^R_2(0)=\phi_2^L(0)$. The boundary condition changing operator from $N \otimes N$ (UV) to $A \otimes N$ (IR) thus corresponds to changing the boundary condition of $\phi_1$ from Neumann to Dirichlet. It thus has dimension $h_{\rm BCC} = \frac{1}{16}$ as in the single channel case. Therefore, we expect the large time behavior of the Loschmidt echo to remain the same as in the non-interacting spinless case,  Eq. (\ref{eqExpG}). In the absence of spin rotation symmetry, a similar calculation leads to the same conclusion (see appendix D in Ref.~\cite{AffleckGiuliano} for a related calculation in a different context).

\subsection{Multichannel case}
\label{Secmultichannel}

We now consider the situation when several channels $\psi_i$ are present in the lead, neglecting interactions. In the single channel case, Andreev reflection is only possible through the Majorana zero mode, since the superconducting term $ \psi \partial_x \psi (0)$ is irrelevant in the renormalization group sense (see Sec.~\ref{SecTopTrivial}). With several channels present, terms like $ \psi_i \psi_j (0)$ with $i\neq j$ are allowed though, representing standard Cooper pair creation/annihilation processes. These are marginal and have to be discussed. After a rotation of the lead channels, ${\tilde{\psi}}_i=\sum_{j}\tilde{U}_{ij} \psi_j$, only a single lead mode $\tilde{\psi}_0$ couples to the Majorana zero mode and the terms of interest generated at the boundary can be written as
\begin{align}
i \kappa \gamma (\tilde{\psi}_0^\dag+\tilde{\psi}_0)(0)+ \sum_{i,j}\lambda_{ij} \tilde{\psi}_i \tilde{\psi}_j (0) + {\rm h.c.},
\label{eq:multi}
\end{align}
where we have assumed the gap to be large such that terms as discussed in Eq.~(\ref{eqNonTopo}) can be neglected. 

The above situation arises {\it e.g.} in the typical semiconductor wire setup for Majorana zero modes~\cite{Oreg,Lutchyn,Delft}, where one channel of the superconducting wire is topological and has essentially a p-wave type gap, while the remaining channels carry the ordinary superconducting correlations inherited from the proximity coupled s-wave superconductor.

The first term in Eq. (\ref{eq:multi}) contributes $-1/4$ to the decay exponent of the Loschmidt echo at large times, in analogy to the single channel case. The remaining boundary terms will in general modify this result, just in the way they would hide the zero-bias conductance peak in a transport experiment. Fortunately the Andreev reflection process underlying these terms relies on the transport of two electrons between lead and superconductor, and is therefore strongly suppressed by a tunneling barrier between the two subsystems, $\lambda_{ij} \sim J'\vphantom{J}^2$, whereas  $\kappa \sim J'$. Since the boundary terms $\lambda_{ij} \tilde{\psi}_i \tilde{\psi}_j (0)$ are marginal, the corrections to the exponent of the Loschmidt echo should depend continuously on $\lambda_{ij} $, and vanish as $ \left| \lambda_{ij}\right| \to 0$. We expect these corrections to scale as $\sim \left| \lambda_{ij}\right|^2$, so that overall the Loschmidt echo will be modified to $t^{-1/4 +{\cal O}((J^\prime)^4)}$ with corrections that are strongly suppressed by a weak barrier. Therefore, in the tunneling regime, formula Eq.~(\ref{eqExpG}) is again recovered up to a reasonable accuracy.

\section{Experimental considerations and quantum dot setup}

\label{secExpQD}

\subsection{Experimental considerations}

At this point, the Loschmidt echo may appear to the reader as a purely theoretical quantity that would be hard to access experimentally. There is however an increasing number of proposals to measure the Loschmidt echo in various local and global quantum quenches -- generalizing the well-known X-ray edge setup~\cite{XRay1,XRay2}, using mostly quantum dot optical absorption~\cite{QuenchKondo0,QuenchKondo1,QuenchKondo3,VM_QSH} and Ramsey interferometry techniques~\cite{DemlerOC,GlobalLL}.

The most promising setup in our context is based on an experiment recently realized to measure post-quench Kondo correlations induced by optical transitions on a quantum dot~\cite{Latta}: when a photon is absorbed by a dot electron, the sudden change caused in the electronic structure can be understood as a local quench between a Fermi reservoir and an effective Kondo impurity. It turns out that the absorption spectrum is essentially the Fourier transform of the Loschmidt echo~\cite{QuenchKondo1}. This yields an edge singularity $A(\omega) \sim \theta (\omega - \omega_0)  (\omega - \omega_0) ^{\alpha/2-1}$ in the low-energy absorption spectrum, with $\alpha$ the exponent that characterizes the power-law decay of the Loschmidt echo at large times. 

The inclusion of a quantum dot between a metallic lead and a topological superconductor could be used in a similar way to induce a quantum quench involving a Majorana zero mode. Quantum dots can naturally be incorporated in most of the normal metal--topological superconductor setups that are currently pursued, even though the actual experimental realization may be challenging. Here we discuss two of the most important cases.
 
The first one is the archetypical Majorana setup based on a spin-orbit coupled semiconducting wire that is proximity coupled to an s-wave superconductor and subjected to a parallel magnetic field~\cite{Oreg, Lutchyn}. When a normal lead is connected to one of its two ends, gate electrodes underneath define tunnel barriers that can create a quantum dot directly at the junction~\cite{Delft}.
The second type of setup is based on an experimentally established topological phase -- the quantum spin Hall effect~\cite{QSH}. A pair of counter-propagating edge states exists at the boundary of this two-dimensional system~\cite{Kane, Bernevig}, which can be turned into a one-dimensional topological superconductor when coupled to an ordinary s-wave superconductor~\cite{Fu}. Uncoupled parts of the edge can serve as a lead, in which tunnel barriers -- and thus a quantum dot -- can be created by depositing ferromagnetic insulators on the edge. 

The actual experimental realization may be challenging though. Consider for example the semiconductor wire setup. The estimated topological gap that can be achieved e.g. for InSb is of the order of $\sim 1{\rm K}$. We require the tunnel broadening ($T^*$) of the Majorana to be considerably smaller , say $0.2{\rm K}$. At the same time, the temperature should again be considerably smaller than the width of the Majorana peak, say $50 {\rm mK}$. This is challenging but generally within reach of current experiments. The temperature requirement might be hard to achieve though if the (weak) radiation heats the sample too fast for the cooling rate. 
 Just as in a transport setting, the topological superconductor further needs to be long enough so that the splitting of the two Majoranas at opposite ends forms the smallest energy scale, $\delta_M\ll 0.2{\rm K}$. 
In addition a large control over the quantum dot is required: the level spacing should be larger than the bandwidths of superconductor and metal so that only the quantum dot absorbs light and the quench is thus really local. At the same time, the tunnel couplings have to be adjusted suitably. For more experimental considerations, we refer the reader to Ref.~\cite{Latta}.

Knowing that a quantum dot setting allows for a measurement of the Loschmidt echo, it is natural to ask whether and how the additional presence of a quantum dot alters the Loschmidt echo in the first place. Without a Majorana zero mode, it is known that the Kondo effect dominates the dynamics~\cite{QuenchKondo1,Latta}, while with a Majorana mode, the Kondo effect competes with the Majorana coupling. It turns out that the Majorana always ``wins''~\cite{InterplayKondoMajo} at low energy/large time, and we will argue in the following that the long time behavior of the Loschmidt echo is given again by Eq.~(\ref{eqExpG}). 

\subsection{Quench dynamics with quantum dot: non-interacting toy model}

\begin{figure}
\centering
    \includegraphics[scale=.25]
    {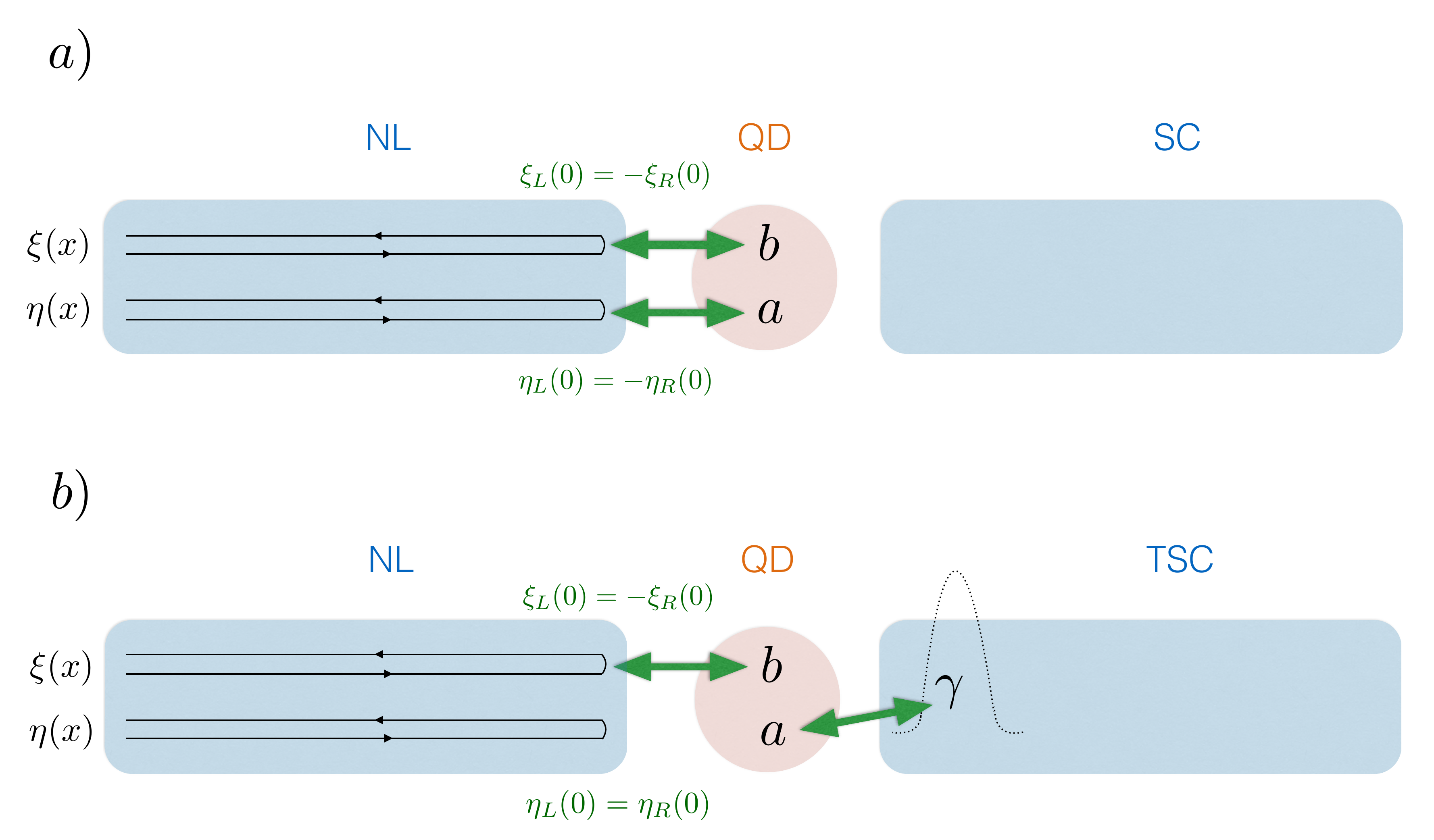}
    \caption{Cartoon representation of the low energy fixed points (large time behavior) for a simplified spinless NL-QD-(T)SC junction with a non-interacting lead. When the superconductor is in a topological phase, the majorana zero-mode at its edge becomes hybridized with ``half'' of the quantum dot, by decomposing the dot fermion as $d=(a +i b)/2$. As a result, only one of the two Majorana channels in the normal lead $\psi(x)=(\xi(x)+i \eta(x))/2$ experiences a $\pi/2$ phase shift corresponding to  Andreev reflection $\psi^\dagger_R(0)=\psi_L(0)$ at the junction. This has crucial consequences on the dynamics after a local quantum quench of the tunneling between the dot and the normal lead.}
\label{Fig1}
\end{figure}

It turns out that the influence of the quantum dot can -- to a large extent -- already be understood from a simple model without interactions on lead or dot. It is a generalization of the low energy Hamiltonian in Eq.~(\ref{FFmajorana}) that includes an extra quantum dot (QD) level $d$ with energy $\epsilon_d$,
\begin{align}
H&= - i \int {\rm d} x\; \psi^\dagger \partial_x \psi + \epsilon_d d^\dagger d + \lambda_1 \left[ \psi^\dagger(0) d + d^\dagger \psi(0)\right] \notag \\ & + i \lambda_2 \gamma \left( d^\dagger + d\right),
\label{eqRLMMajo}
\end{align}
coupled to both the normal lead and the Majorana zero mode. Following~\cite{QuenchKondo0,QuenchKondo1,QuenchKondo3}, an optical transition on the dot can be understood as an effective quantum quench of the tunneling $\lambda_1$. 

While the above non-interacting model may seem artificial in its form, eq.~\eqref{eqRLMMajo} for $\lambda_2=0$ (also known as the resonant level model) exhibits a sort of ``Kondo physics'' at low energy -- it corresponds to the so-called Toulouse point of the anisotropic Kondo problem~\cite{Kondo}. It is thus the simplest example to study the interplay between Kondo and Majorana physics.

In order to analyze the long-time post-quench dynamics, it is very convenient to write both the lead and the dot excitations in terms of Majorana operators, $\psi=(\xi+i \eta)/2$, $d=(a+i b)/2$. Then the problem reduces to two independent Majorana problems $H=H_\xi + H_\eta$
\begin{align}
H_\xi &= - \frac{i}{4} \int {\rm d} x \xi \partial_x \xi + i \frac{\lambda_1}{2} \xi(0) b, \notag \\
H_\eta &= - \frac{i}{4} \int {\rm d} x \eta \partial_x \eta - i \frac{\lambda_1}{2} \eta(0) a + i \lambda_2 \gamma a,
\label{eqNonIntKondoMaj}
\end{align}
where we set $\epsilon_d=0$ for simplicity. The scattering matrix for $\xi(x)$ is formally equivalent to~\eqref{eqScattKondo}, leading to $\xi(0^+)=-\xi(0^-)$ as low energy boundary condition. For the field $\eta(x)$, the situation is slightly more complicated. When the coupling to the Majorana vanishes, $\lambda_2=0$, the solution is essentially the same as for $\xi(x)$ and we obtain $\eta(0^+)=-\eta(0^-)$ at low energy as well. 
If on the other hand, $\lambda_2\neq 0$, the scattering matrix for $\eta(x)$ is modified to
\begin{equation}
S_\eta(\omega) = \frac{i \omega - \frac{\lambda_1^2}{2} \left( \frac{\omega^2}{\omega^2+ 4 \lambda_2^2}\right)}{i \omega + \frac{\lambda_1^2}{2} \left( \frac{\omega^2}{\omega^2+ 4 \lambda_2^2}\right)},
\end{equation}
so that now $\eta(0^+)=\eta(0^-)$ at low energy.

In terms of the complex fermion $\psi(x)$, the boundary conditions at low energy ($\omega\rightarrow 0$) are
\begin{align}
\psi(0^+)= \mathrm{e}^{2 i \delta} \psi(0^-)\hspace{20pt}{\rm for\; \lambda_2=0}\label{eq:bound1},\\
 \psi^\dagger(0^+)=\psi(0^-)\hspace{20pt}{\rm for\; \lambda_2\neq0}\label{eq:bound2},
\end{align}
with $\delta=\frac{\pi}{2}$ for $\epsilon_d=0$ (in the general case, $\mathrm{e}^{2 i \delta} =(i \epsilon_d - \frac{\lambda_1^2}{2})/(i \epsilon_d + \frac{\lambda_1^2}{2})$). This corresponds to a Kondo-type boundary condition (no Majorana) and an Andreev boundary condition (with Majorana), respectively. We can therefore observe in this very simple example how the Majorana coupling ``wins'' at low energy over the Kondo coupling. 

After a quench in $\lambda_1$, this leads to a large time behavior of the Loschmidt echo given by
\begin{align}
{\cal L}_{\lambda_2=0}(t) &\sim t^{-2 (\delta/\pi)^2},\label{eq:LKondo}\\
{\cal L}_{\lambda_2\neq0}(t) &\sim t^{-1/4}.\label{eq:LMaj}
\end{align}
Note that a phase like $\mathrm{e}^{2 i \delta}$ as it occurs in Eq. (\ref{eq:bound1}) can in principle also occur in Eq. (\ref{eq:bound2}). But in the latter case it can be readily absorbed by a canonical transformation $\psi \mapsto \mathrm{e}^{-i \delta} \psi$, $ \psi^\dagger \mapsto \mathrm{e}^{i \delta} \psi^\dagger$.

In the renormalization group picture, this simple non-interacting model provides a very intuitive explanation of the IR fixed point: indeed, one can easily see from~\eqref{eqNonIntKondoMaj} that if $\lambda_2=0$, the ``Kondo'' fixed point $\lambda_1 \to \infty$ enforces $\eta(0)=\frac{\eta(0^+)+\eta(0^-)}{2}=0$ and $\xi(0)=\frac{\xi(0^+)+\xi(0^-)}{2}=0$, so that $\psi(0^+)=-\psi(0^-)$. If $\lambda_2\neq 0$, as we have seen above, the Majorana coupling term is more relevant in the RG sense and prevails over the $\lambda_1$ coupling at low energy. Introducing a new fermion operator $\tilde{d}=\frac{\gamma+i a}{2}$ so that $i \lambda_2 \gamma a = \lambda_2( 2 \tilde{d}^\dagger \tilde{d}-1)$, the fixed point $\lambda_2 \to \infty $ will polarize the effective fermion $\tilde{d}$ and enforce $\langle \tilde{d}^\dagger \tilde{d} \rangle= 0$ at low energy. Therefore, the IR boundary condition satisfied by the Majorana field $\eta(x)$ remains free  $\eta(0^+)=\eta(0^-)$, implying $\psi^\dagger(0^+)=\psi(0^-)$. This simple picture is summarized in Fig.~\ref{Fig1}. This suggests that only the parity of the number of Majorana fermions at the boundary matters: the quantum dot can be thought of as adding two majorana fermions $a,b$ at the boundary, thus leaving the total parity (odd) unchanged. 

\subsection{Quench dynamics with quantum dot: general case}

To understand how this physical picture translates to a more realistic setup, we consider an Anderson impurity tunnel-coupled to a (spinful) normal lead and a Majorana, say polarized along the spin $\uparrow$ (assuming spin-rotation symmetry for simplicity) 
\begin{align}
H&=H_{\rm L} + U n_{\uparrow} n_{\downarrow} + \epsilon_d \left( n_{\uparrow} +n_{\downarrow} \right) \notag \\ & + \lambda_1 \sum_\sigma \left(\psi^\dagger_\sigma(0) d_\sigma +{\rm h.c.} \right) + i \lambda_2 \gamma \left( d^\dagger_\uparrow +  d_\uparrow \right),
\end{align}
where $H_{\rm L}$ refers to the interacting lead Hamiltonian in Eq. (\ref{eq:spinfullead}) and we assume that the interaction energy $U$ is the dominant energy scale (Kondo regime).

If $\lambda_2 =0$, the system is known to flow to a Kondo low energy fixed point $\psi_\sigma(0^+)= \mathrm{e}^{2 i \delta_\sigma}\psi_\sigma(0^-)$ -- after unfolding the lead, where $\delta_\sigma$ is the phase shift for the spin $\sigma$ electrons.  After a quench in $\lambda_1$, this leads to a large time behavior of the Loschmidt echo similar to Eq. (\ref{eq:LKondo}), with two independent contributions coming from the two spin channels, where the phase shifts $\delta_\uparrow$, $\delta_\downarrow$ can be tuned by applying a magnetic field on the quantum impurity. This is the case that was studied experimentally in Ref.~\cite{Latta}. At the particle hole symmetric point, $\delta_\downarrow = \delta_\uparrow = \frac{\pi}{2}$ so that ${\cal L}(t) \sim t^{-1}$.
 
Once $\lambda_2 \neq 0$, the Kondo and Majorana couplings compete. It was recently conjectured~\cite{InterplayKondoMajo}  for this model, based on a perturbative RG analysis and on DMRG simulations, that the Majorana coupling dominates at low energy and that the boundary condition characterizing the IR physics is $A \otimes N$ (corresponding to Andreev reflection for the spin up channel, while the down channel experiences normal reflection), just as in the absence of the quantum dot.
In Appendix~\ref{AppKondoStab} we give further arguments why this is indeed the case. The operator changing boundary conditions from $N \otimes N$ to $A \otimes N$ again has dimension $h_{\rm BCC}=\frac{1}{16}$, so the Loschmidt echo still behaves as~\eqref{eq:LMaj} in this case. 

However, we note that we have implicitly assumed that the spin down channel subject to normal reflection at low energy actually satisfies $\psi_\downarrow(0^+)=\psi_\downarrow(0^-)$, without any phase shift. It is hard to argue that this is true in general in the context of the Kondo effect -- even though qualitative arguments tend to indicate that this is indeed the case~\cite{InterplayKondoMajo}, and it might be that the spin down channel in the boundary condition $A \otimes N$ feels some phase shift due to the quantum dot. In this case, the Loschmidt echo would have two contributions ${\cal L}(t)\sim  t^{-1/4} t^{-2 (\delta_\downarrow/\pi)^2}$. Nevertheless, we emphasize that the phase shift $\delta_\downarrow$ would be tunable by applying a magnetic field on the dot, and that it could be independently measured in the absence of the Majorana coupling. Hence, the universal Majorana signature $\sim t^{-1/4}$ could still be extracted in this case.

\section{Conclusion}

To conclude, we have argued that the presence of Majorana bound states at the edge of topological superconductors can be probed using local quantum quenches. The Majorana zero mode acts as a quantum impurity thereby inducing a sort of Anderson orthogonality catastrophe with a universal, ``quantized'' exponent for the wave function overlap ${\cal{L}}(t)= \left| \Braket{\psi_0 | \psi(t)} \right|^2 \sim t^{-1/4}$. This exponent was shown to be as robust as the zero-bias anomaly in the tunnel conductance against  interactions in the normal lead and additional channels. The most promising setup to measure this exponent involves optical absorption of a quantum dot between the lead and the superconductor, inducing an effective quantum quench. This robustness can be traced back to the irrelevance of the phase degree of freedom in the boundary condition $\psi^\dagger(0^+)=\psi(0^-)$ since these can be absorbed by a simple $U(1)$ transformation. In contrast the phase $\delta$ in the boundary condition $\psi(0^+)={\rm e}^{2i \delta}\psi(0^-)$ strongly influences the dynamics and generally depends on many non-universal details.  An interesting problem for future investigation would be to check whether disorder effects complicate the distinction between topological and non-topological case in the same way as in the typical transport setting~\cite{DisorderedJan,DisorderedAltland,DisorderedDrew}.

We finally mention that similar quantum quenches could be used to probe edge modes in other kinds of (symmetry protected) topological 1D systems. One could for example consider an antiferromagnetic spin-1 chain, which is gapped~\cite{Haldane} and known to host fractionalized spin-$\frac{1}{2}$ edge excitations protected by spin rotation symmetry~\cite{AKLT}. Suddenly coupling these excitations to a Fermi reservoir should lead to Kondo correlations in the time dynamics similar to those observed in Ref.~\cite{Latta}.

Moreover, our results also apply when the lead is replaced by a Majorana edge state, e.g. at the edge of a $p+i p$ superconductor~\cite{ReadGreen}, or the neutral sector of the $\nu=\frac{5}{2}$ Moore-Read state~\cite{MooreRead}. In this setting, Majorana zero modes naturally appear bound to vortex cores in the bulk and the quench could  be induced by a gate that forces the edge state inwards, coupling it to one such zero mode~\cite{Annals}.

To see how this concept generalizes to other (non-abelian) topological states, consider for example a quench involving a $\mathbb{Z}_3$ parafermionic zero mode~\cite{Fendley} coupled to a gapless $\mathbb{Z}_3$ parafermionic theory. For the former an experimental realization was proposed in Ref.~\cite{Para, PRX}, while the latter appears both at the edge of the Read-Rezayi $\nu=\frac{13}{5}$ state modulo the charged boson~\cite{ReadRezayi}, and at the edge of the ``Fibonacci superconducting phase'' introduced in Ref.~\cite{PRX}. 

We expect the $\mathbb{Z}_3$ parafermionic edge zero mode to act as a boundary magnetic field perturbation, with dimension $\Delta=\frac{2}{5}$, for the gapless $\mathbb{Z}_3$ parafermionic theory (describing the critical point of the 2D $Q=3$-state Potts model, just like the gapless Majorana theory describes the critical point of the Ising model). The conformally invariant boundary conditions of the Potts model are well-known and classified~\cite{Potts3}, and the boundary condition changing operator from free to fixed in the $Q=3$ Potts model has dimension $h_{\rm BCC}=\frac{1}{8}$~\footnote{Note that there is also the possibility of a mixed boundary condition for non-generic values of the magnetic field.}. This would lead to a Loschmidt echo decaying as ${\cal{L}}(t) \sim t^{-1/2}$. We expect that other kinds of edge zero modes, including higher-order parafermions, could be analyzed from this quench perspective in the same way.  The general lesson appears to be that quantum quenches give surprisingly precise and robust information about localized topological excitations.

\
\

{\it Acknowledgments.} 
We thank R. Lutchyn, S. Parameswaran, A.C. Potter and H. Saleur for insightful discussions. RV also wishes to thank D. Kennes and V. Meden for collaborations on related matters. This work was supported by the Quantum Materials program of LBNL (R.V.), NSF DMR-1206515 and the Simons Foundation (J.E.M.), the Netherlands Organization for Scientific Research NWO (J.P.D.) and the German Academic Exchange Service DAAD (J.P.D.).

\appendix

\section{Asymptotic behavior of the Loschmidt echo from BCFT}
\label{AppLoschBCC}

\begin{figure}
\centering
    \includegraphics[width=\columnwidth]
    {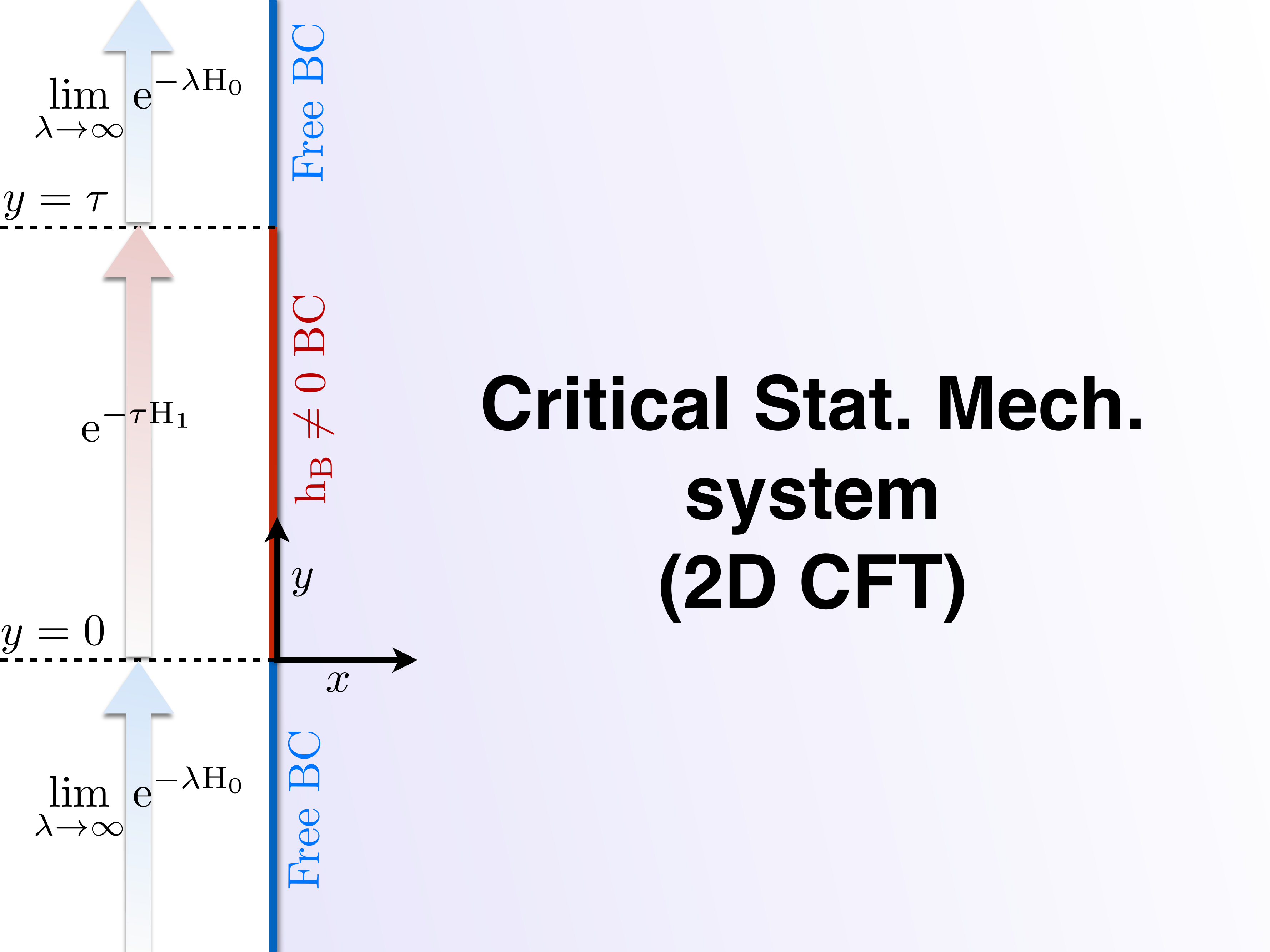}
    \caption{Geometry of the partition function $Z_{0,1}(\tau)$ of the 2D statistical mechanics problem in the half-plane $x \in [0,  \infty)$, $y \in (-\infty,  \infty)$ associated with the analytic continuation ${\cal L} (t=- i \tau ) $ of the Loschmidt echo in imaginary time. The quantum quench then amounts to changing the boundary condition at $x=0$ for $y \in [0,  \tau]$. }
\label{FigLoschGeom}
\end{figure}

In this appendix, we provide more details on the calculation of the asymptotic behavior of the Loschmidt echo ${\cal L}(t) \sim t^{-\alpha}$ from Boundary Conformal Field Theory (BCFT). Let us consider a 1D quantum system on the half-line $x\in[0, \infty)$ in a pure state $\Ket{\psi_0}$, which is the ground state of a gapless Hamiltonian $H_0$ with energy $E_0$, given a free boundary condition at $x=0$. At time $t=0$, we abruptly change the Hamiltonian to $H_1$, where $H_1$ differs from $H_0$ by a boundary term acting at $x=0$. This can be interpreted as adding a sort of boundary magnetic field $h_B\neq 0$ to  $H_0$ (in our context, this corresponds to adding the Majorana coupling). The Loschmidt echo then reads ${\cal L}(t)=\left| \Braket{\psi_0 \left| {\rm e}^{-it H_1}\right| \psi_0} \right|^2$. To proceed, we first perform a Wick rotation $t=- i \tau $, and express $\Ket{\psi_0}$ using the relation
\begin{equation}
\Ket{\psi_0} = \lim_{\ell \to \infty} \frac{{\rm e}^{-\ell (H_0-E_0) }\Ket{\alpha}}{\Braket{\psi_0 | \alpha}},
\end{equation}
valid for any generic state $\Ket{\alpha}$ not orthogonal to $\Ket{\psi_0}$. One can therefore rewrite the Loschmidt echo as 
\begin{equation}
{\cal L} (t=- i \tau ) \propto \left| Z_{0,1}(\tau) \right|^2, 
\end{equation}
where $Z_{0,1}(\tau)= \lim_{\ell \to \infty} \Braket{\alpha \right| {\rm e}^{-\ell H_0}  {\rm e}^{-\tau H_1} {\rm e}^{-\ell H_0}\ \left| \alpha}$ can be interpreted as the partition function of a 2D classical statistical mechanics problem in the half-plane $x \in [0,  \infty)$, $y \in (-\infty,  \infty)$, critical in the bulk ($H_0$ being gapless), with $y$ the imaginary time direction.  
In that language, the imaginary time evolution operators $T_{0,1}(\lambda)= {\rm e}^{-\lambda H_{0,1}}$ now correspond to transfer matrices with different boundary conditions at $x=0$ (free for $T_0$, and ``boundary magnetic field $h_B$'' for $T_1$). The quantum quench thus amounts to changing the boundary condition at $x=0$ for $y \in [0,  \tau]$ in a two-dimensional conformal field theory. The corresponding geometry is sketched in Fig.~\ref{FigLoschGeom}. For a generic boundary perturbation in $H_1$, this partition function remains very complicated, but for large $\tau$, one expects the boundary condition to flow to a conformally invariant one -- roughly speaking, the ``boundary magnetic field'' $h_B$ flows to infinity. The geometry of the partition function $Z_{0,1}(\tau)$ then coincides precisely with the definition of the two-point function of a Boundary Condition Changing (BCC) operator $\phi_{0 \to 1}$ in the BCFT language~\cite{Cardy}. For large $\tau$ -- compared to all the crossover scales in the problem -- we can thus write 
\begin{equation}
Z_{0,1} (\tau) \sim \langle  \phi_{1 \to 0} (x=0,y=\tau) \phi_{0 \to 1} (x=0,y=0) \rangle.   
\end{equation}
We thus expect $Z_{0,1} (\tau) \sim \tau^{-2 h_{\rm BCC}}$, where $h_{\rm BCC}$ is the scaling dimension of the operator $\phi_{0 \to 1}$. Going back to the Loschmidt echo in real time, this yields $\alpha= 4 h_{\rm BCC}$ as claimed in the main text.

To illustrate this construction on a concrete example, let us go back to the single-channel non-interacting case. As discussed in Sec.~\ref{secKitaev}, the Hamiltonian $H_0$ can be written as two independent massless Majorana theories. The corresponding 2D statistical mechanics problem is thus given by two decoupled copies of the critical Ising model, with total central charge $c=1=\frac{1}{2}+\frac{1}{2}$, and the quantum quench corresponds to adding a finite boundary magnetic field to one of these copies  for $y \in [0,  \tau]$, with free boundary conditions elsewhere. The other copy of the Ising model does not couple to the Majorana zero mode and can therefore be dropped. A finite magnetic field at the boundary of the Ising model is not a conformally invariant boundary condition, but it is well known to flow to a ``fixed'' boundary condition, and the corresponding critical exponent is given by the spin operator $h_{\rm BCC}=\frac{1}{16}$~\cite{Cardy}.

\section{Free Fermions Numerics}

\label{AppNumerics}

In this appendix, we provide technical details on the numerical evaluation of the Loschmidt echo for the non-interacting (quadratic) system described in section~\ref{secKitaev}. The whole system can then be solved using Bogoliubov de Gennes equations. Let us start with the Hamiltonian $H_q$ after the quench, which we rewrite as 
\begin{align}
H_q=
\begin{pmatrix}
\bf{c}^\dag \  \bf{c}
\end{pmatrix}
\mathcal{H}_{\rm BdG}
\begin{pmatrix}
\bf{c}\\
\bf{c}^\dag
\end{pmatrix}-E_0
\end{align}
with $\mathbf{c}=(c_{-L+1},\ldots,c_0,c_1,\ldots c_L)$ a vector of destruction operators. Here we have replaced the notation of section~\ref{secKitaev} using $c_{-i+1}=f_i$. By adjusting the scalar $E_0$, the $4L\times4L$ matrix $\mathcal{H}_{BdG}$ can be chosen hermitian and to obey Particle Hole Symmetry (PHS), $\mathcal{H}_{BdG}=-\tau_x \mathcal{H}^*_{BdG}\tau_x$. We now diagonalise the Bogoliubov de Gennes Hamiltonian $\mathcal{H}_{BdG}=V_q \Lambda V_q^\dag$, with $\Lambda$ a diagonal matrix of eigen energies. Due to PHS we can choose the eigenvectors and their order such that $V_q$ obeys $\tau_x V_q^* \tau_x=V_q$, implying the following energy ordering: $\Lambda={\rm diag } (-E_1,\ldots, -E_{2L},E_1,\ldots,E_{2L})$ with all $E_i>0$. Then, $\bf{q}^\dag$ with
\begin{align}
\begin{pmatrix}
\bf{q}^\dag\\
\bf{q}
\end{pmatrix}=
V_q^\dag
\begin{pmatrix}
\bf{c}\\
\bf{c}^\dag
\end{pmatrix},
\end{align}
is the vector of creation operators of the systems excitations, such that
\begin{align}
 H_q=\sum_{i=1}^{2L} 2 E_i q_i^\dag  q_i -\sum_{i=0}^{2L} E_i.
\end{align}
For the evaluation of the Loschmidt echo we will also need the eigenvectors $V_d$ for the situation before the quench ($J'=0$), which can be obtained in the same way. Then, the excitations $d^\dag_i$ of the initial Hamiltonian can be expressed in terms of the excitations $q^\dag_i$ of the quenched Hamiltonian
\begin{align}
\begin{pmatrix}
\bf{d}^\dag\\
\bf{d}
\end{pmatrix}=V_d^\dag V_q
\begin{pmatrix}
\bf{q}^\dag\\
\bf{q}
\end{pmatrix}=
\begin{pmatrix}
A & B\\
B^* & A^*
\end{pmatrix}
\begin{pmatrix}
\bf{q}^\dag\\
\bf{q}
\end{pmatrix}.
\label{eq:qtod}
\end{align}

The Loschmidt echo is given by the overlap of the initial many-body groundstate $|\psi_d\rangle$ with the actual state of the system at time $t$, after propagation with the Hamiltonian $H_g$ that governs the dynamics after the quench,
\begin{align}
\mathcal{L}(t)&=\left|\langle \psi_d |e^{-iH_q t}|\psi_d \rangle\right|^2.
\end{align}
As is discussed in detail in Appendix A of Ref. \cite{Ste11}, the Loschmidt echo can be expressed in terms of determinants of matrices,
\begin{align}
\mathcal{L}(t)&=\frac{\mathcal{N}(t)}{\mathcal{N}(0)}=\frac{|\det (N^{(t)}+M)|^2}{|\det (N^{(0)}+M)|^2},
\label{eq:Loschmidt}
\end{align}
 with $M_{jl}=\langle \psi_q|d^\dag_j d^\dag_l |\psi_q\rangle$, and $N_{jl}^{(t)}=\langle \psi_q| d^\dag_j d_l(t)|\psi_q\rangle$.
Here, $|\psi_q\rangle$ denotes the gound state of the system after the quench. 
Using Eq. (\ref{eq:qtod}) and the fact that all destruction operator $q_i$ negate the groundstate $|\psi_q\rangle$, we obtain
\begin{align}
M_{jl}&=\sum_k b_{jk} a_{lk}= (B A^T)_{jl},\\
N_{jl}^{(t)}&=\sum_k e^{-iE_k t} b_{jk} b_{lk}^*= (B D_t B^\dag)_{jl},
\end{align}
in terms of the matrices $A$, $B$ and their elements $a_{ij}$, $b_{ij}$ as well as the diagonal matrix $D_t={\rm diag } (e^{-iE_k t})$.
Inserting these expressions into Eq. (\ref{eq:Loschmidt}) and crossing out a factor $\det B$, we obtain as final expression
\begin{align}
\mathcal{L}(t)&=\frac{|\det (A^T+D_tB^\dag)|^2}{|\det (A^T+B^\dag)|^2}.
\end{align}
Note that this last simplification is crucial for the numerical stability, since $\det B$ can become vanishingly small for large systems. 
Furthermore during the numerical evaluation great care is required to achieve the correct ordering of the eigenstates and to guarantee their symmetry property $\tau_x V^* \tau_x=V$.

\section{Stability of the IR Kondo fixed point and boundary entropy counting}

\label{AppKondoStab}

In section~\ref{secExpQD}, we argued that the large time behavior after a quench of the tunneling between a quantum dot and a normal lead was dominated by Majorana physics if the quantum dot is also tunnel-coupled to a topological superconductor (NL-QD-TSC junction).

Another way to state this result is to claim that the Kondo fixed point is unstable upon coupling to a single Majorana fermion. Proving this is actually almost trivial when the lead is non-interacting. Even in strongly interacting quantum impurity problems that flow to a Kondo fixed point -- say the Anderson model in the Kondo regime, the (conformally invariant) field theory describing the low energy Kondo point is very simple: restricting ourselves to a single channel, it consists of a massless chiral fermion (for a non-interacting lead) satisfying the boundary condition $\psi(0^+)=\mathrm{e}^{2 i \delta}\psi(0^-)$, with $\delta$ a phase shift. This boundary condition can be very conveniently implemented through a scattering potential $V(\delta)$ at $x=0$ so that the Hamiltonian at the fixed point reads
\begin{equation}
H^{\star}= - i v_F \int {\rm d} x \psi^\dagger \left[\partial_x + i V(\delta) \delta(x) \right] \psi,
\end{equation}
where $V(\delta) \simeq 2 \delta + \dots$ The precise relation between $V$ and $\delta$ depends on the regularization of the Dirac delta function but this is irrelevant for our discussion. The only thing that matters is that the scattering potential can be interpreted as a gauge potential that can be gauged away by enforcing $\psi(0^+)=\mathrm{e}^{2 i \delta}\psi(0^-)$. Let us now couple the system to a single Majorana fermion $\gamma$: the most relevant perturbation then corresponds to the perturbed Hamiltonian 
\begin{equation}
H= H^{\star} + \lambda i \gamma \left( \psi^\dagger +\psi \right)(0),
\end{equation}
which has dimension $\Delta=\frac{1}{2}$ and is therefore relevant. This (trivially) proves that the Kondo fixed point unstable when coupled to a Majorana mode. Moreover, since the perturbed Hamiltonian is still quadratic, one can readily compute the scattering matrix~\eqref{eqDefSMatrix} to show explicitly that the perturbation drives the boundary condition from Kondo $\psi(0^+)=\mathrm{e}^{2 i \delta}\psi(0^-)$ to Andreev $\psi^\dagger(0^+)=\psi(0^-)$. 

When the lead is interacting, this remains true whenever the perturbation $\lambda i \gamma \left( \psi^\dagger +\psi \right)(0)$ is relevant, that is, whenever its scaling dimension $\Delta$ satisfies $\Delta<1$. So as long as the Majorana fixed point is stable in the first place~\cite{AffleckGiuliano}, the quantum dot does not modify the nature of the low energy fixed point. Going back to the original problem of coupling a spinful Luttinger liquid to a quantum dot and a Majorana zero mode, one can also see that the $A \otimes N$ fixed point is more stable than the Kondo fixed point using boundary entropy counting. The entropy drop going from the UV fixed point to the Kondo fixed point satisfies
\begin{equation}
\Delta S = \ln g_{\rm Kondo} - \ln g_{\rm UV}=-\ln 2,
\end{equation}
since the impurity with two degrees of freedom that is free at high energy becomes hybridized (screened) with the wire at the Kondo fixed point. On the other hand, the $g$ factor of the $A \otimes N$ fixed point satisfies 
\begin{equation}
\Delta S = \ln g_{A \otimes N}- \ln g_{\rm UV} =-\ln \frac{2}{\sqrt{\Delta}},
\end{equation}
since the quantum dot is fully polarized at the $A \otimes N$ fixed point, and that the energy drop associated with a change from Neumann to Dirichlet boundary conditions (in the bosonization language) is $ \frac{1}{2}\ln \Delta < 0$ since $\Delta<1$. A RG flow from Kondo to $A \otimes N$ can only exist if the associated $g$ factor decreases along the flow, that is
\begin{equation}
\frac{g_{\rm Kondo}}{g_{A \otimes N}}=\frac{1}{\sqrt{\Delta}} > 1.
\end{equation}
We therefore recover that as long as $\Delta < 1$, the  $A \otimes N$ fixed point is more stable than the Kondo fixed point.


\begin{thebibliography}{99}
\bibitem{andersonorthogonality} P.W. Anderson, {\sl Infrared Catastrophe in Fermi Gases with Local Scattering Potentials}, Phys. Rev. Lett. {\bf 18}, 1049--1051 (1967).


\bibitem{XRay1} G. D. Mahan, {\sl Excitons in Metals: Infinite Hole Mass}, Phys. Rev. {\bf 163}, 612 (1967).

\bibitem{XRay2} P. Nozi\`urea and C. T. de Dominicis, {\sl Singularities in the X-Ray Absorption and Emission of Metals. III. One-Body Theory Exact Solution}, Phys. Rev. {\bf 178},  1097 (1969).

\bibitem{Calabrese:2006} P. Calabrese and J. Cardy, {\sl Time dependence of correlation functions following a quantum quench}, Phys. Rev. Lett.  {\bf 96}, 136801 (2006).

\bibitem{Lamacraft:2007p841} A. Lamacraft, {\sl Quantum Quenches in a Spinor Condensate}, Phys. Rev. Lett. {\bf 98}, 160404 (2007).

\bibitem{Lamacraft2012177}  A. Lamacraft and J. Moore,  {\sl Chapter 7 - Potential Insights into Nonequilibrium Behavior from Atomic Physics}, Ultracold Bosonic and Fermionic Gases, 177 - 202, Elsevier (2012).


\bibitem{Latta} C. Latta  {\sl et al.}, {\sl Quantum quench of Kondo correlations in optical absorption}, Nature, \textbf{474}, 627 -- 630 (2011).

\bibitem{Caux} J-S. Caux and F.H.L. Essler, {\sl Time Evolution of Local Observables After Quenching to an Integrable Model}, Phys. Rev. Lett. {\bf 110}, 257203 (2013), 


\bibitem{Rigol} M. Rigol, {\sl Quantum Quenches in the Thermodynamic Limit}, Phys. Rev. Lett. {\bf 112}, 170601 (2014).

\bibitem{RevZhang} X.-L. Qi and S.-C. Zhang, {\sl Topological insulators and superconductors}, Rev. Mod. Phys. {\bf 83}, 1057 (2011).

\bibitem{RevAlicea} J. Alicea, {\sl New directions in the pursuit of Majorana fermions in solid state systems}, Rep. Prog. Phys. {\bf 75}, 076501 (2012).

\bibitem{RevBeenakker} C.W.J. Beenakker, {\sl Search for Majorana fermions in superconductors}, Annu. Rev. Con. Mat. Phys. {\bf 4}, 113 (2013).

\bibitem{Kitaev} A. Y. Kitaev, {\sl Unpaired Majorana fermions in quantum wires}, Physics-Uspekhi {\bf 44}, 131 (2001).


\bibitem{Oreg} Y. Oreg, G. Refael, and F. von Oppen, {\sl Helical Liquids and Majorana Bound States in Quantum Wires}, Phys. Rev. Lett. \textbf{105}, 177002 (2010).

\bibitem{Lutchyn} R. M. Lutchyn, J. D. Sau, and S. Das Sarma, {\sl Majorana Fermions and a Topological Phase Transition in Semiconductor-Superconductor Heterostructures}, Phys. Rev. Lett. \textbf{105}, 077001 (2010).

\bibitem{Delft} V. Mourik {\sl et al.}, {\sl Signatures of Majorana Fermions in Hybrid Superconductor-Semiconductor Nanowire Devices}, Science \textbf{336}, 1003 (2012).

\bibitem{Deng} M. T. Deng, C. L. Yu, G. Y. Huang, M. Larsson, P. Caroff, and H. Q. Xu, {\sl Anomalous Zero-Bias Conductance Peak in a NbÐInSb NanowireÐNb Hybrid Device}, Nano Lett. {\bf 12}, 6414 (2012).

\bibitem{Rokh} L. P. Rokhinson, X. Liu, and J. K. Furdyna, {\sl The fractional a.c. Josephson effect in a semiconductor- superconductor nanowire as a signature of Majorana particles}, Nat. Phys. {\bf 8}, 795 (2012).

\bibitem{Das} A. Das, Y. Ronen, Y. Most, Y. Oreg, M. Heiblum, and H. Shtrikman, {\sl Zero-bias peaks and splitting in an Al- InAs nanowire topological superconductor as a signature of Majorana fermions}, Nat. Phys. {\bf 8}, 887 (2012).

\bibitem{Marcus} H. O. H. Churchill, V. Fatemi, K. Grove-Rasmussen, M. T. Deng, P. Caroff, H. Q. Xu, C. M. Marcus, {\sl Superconductor-Nanowire Devices from Tunneling to the Multichannel Regime: Zero-Bias Oscillations and Magnetoconductance Crossover},  Phys. Rev. B {\bf 87}, 241401(R) (2013).

\bibitem{Finck}A. D. K. Finck, D. J. Van Harlingen, P. K. Mohseni, K. Jung, and X. Li, {\sl Anomalous Modulation of a Zero-Bias Peak in a Hybrid Nanowire-Superconductor Device}, Phys. Rev. Lett. {\bf 110}, 126406 (2013).

\bibitem{Franceschi} E.J. H. Lee, X. Jiang, M. Houzet, R. Aguado, C.M. Lieber and S. De Franceschi, {\sl Spin-resolved Andreev levels and parity crossings in hybrid superconductor-semiconductor nanostructures}, Nature Nanotechnology {\bf 9}, 79--84 (2014).

\bibitem{Cardy} John L. Cardy, {\sl Conformal invariance and surface critical behavior}, Nucl. Phys. B {\bf 240},  514--532 (1984), {\sl Effect of boundary-conditions on the operator content of two-dimensional conformally invariant theories}, Nucl. Phys. B {\bf 275},  200--218 (1986), {\sl Boundary conditions, fusion rules and the Verlinde formula}, Nucl. Phys. B {\bf 324},  581--596 (1989).

\bibitem{White} S. R. White, {\sl Density matrix formulation for quantum renormalization groups}, Phys. Rev. Lett. {\bf 69}, 2863 (1992).

\bibitem{Vidal1} G. Vidal, {\sl Efficient Classical Simulation of Slightly Entangled Quantum Computations}, Phys. Rev. Lett. {\bf 91}, 147902 (2003).

\bibitem{Vidal2} G. Vidal, {\sl Classical Simulation of Infinite-Size Quantum Lattice Systems in One Spatial Dimension}, Phys. Rev. Lett. {\bf 98}, 070201 (2007).

\bibitem{Schollwoeck} U. Schollwoeck, {\sl The density-matrix renormalization group in the age of matrix product states}, Annals of Physics {\bf 326}, 96 (2011).

\bibitem{Sengupta} H.-J. Kwon, K. Sengupta and V. M. Yakovenko, {\sl Fractional ac Josephson effect in p- and d-wave superconductors}, Eur. Phys. J. B {\bf 37}, 349--361 (2004).

\bibitem{Bolech} C. J. Bolech and E. Demler, {\sl Observing Majorana bound states in p-wave superconductors using noise measure- ments in tunneling experiments}, Phys. Rev. Lett. {\bf 98}, 237002 (2007).

\bibitem{Nilsson} J. Nilsson, A. R. Akhmerov, and C. W. J. Beenakker, {\sl Splitting of a Cooper pair by a pair of Majorana bound states}, Phys. Rev. Lett. {\bf 101}, 120403 (2008).

\bibitem{LawLee} K. T. Law, Patrick A. Lee, and T. K. Ng, {\sl Majorana Fermion Induced Resonant Andreev Reflection} Phys. Rev. Lett. {\bf 103}, 237001 (2009).

\bibitem{Fu2}  L. Fu, {\sl Electron teleportation via Majorana bound states in a mesoscopic superconductor}, Phys. Rev. Lett. {\bf 104}, 056402 (2010).

\bibitem{FuKane} L. Fu and C. L. Kane, {\sl Josephson current and noise at a superconductor-quantum spin Hall insulator- superconductor junction}, Phys. Rev. B {\bf 79}, 161408(R) (2009).

\bibitem{Wimmer} M. Wimmer, A. R. Akhmerov, J. P. Dahlhaus, and C. W. J. Beenakker, {\sl Quantum point contact as a probe of a topological superconductor}, New J. Phys. {\bf 13}, 053016 (2011).

\bibitem{Akhmerov} A. R. Akhmerov, J. P. Dahlhaus, F. Hassler, M. Wimmer, and C. W. J. Beenakker, {\sl Quantized conductance at the Majorana phase transition in a disordered superconducting wire}, Phys. Rev. Lett. {\bf 106}, 057001 (2011).


\bibitem{MajoranaLuttinger}  L. Fidkowski, J. Alicea, N. H. Lindner, R. M. Lutchyn, and M. P. A. Fisher, {\sl Universal transport signatures of Majorana fermions in superconductor-Luttinger liquid junctions}, Phys. Rev. B {\bf 85}, 245121 (2012).

\bibitem{gFactor} I. Affleck and A.W.W. Ludwig, {\sl Universal noninteger ``ground-state degeneracy'' in critical quantum systems}, Phys. Rev. Lett. {\bf 67}, 161 (1991).

\bibitem{QuenchBCC} R. Vasseur, K. Trinh, S. Haas and H. Saleur,  {\sl Crossover Physics in the Nonequilibrium Dynamics of Quenched Quantum Impurity Systems}, 
Phys. Rev. Lett. {\bf 110}, 240601 (2013).

\bibitem{chargeflow1}  A. Weichselbaum, W. M\"under and J. von Delft, {\sl Anderson orthogonality and the numerical renormalization group}, Phys. Rev. B {\bf 84},  075137 (2011).

\bibitem{chargeflow2} W. M\"under, A. Weichselbaum, M. Goldstein, Y. Gefen and J. von Delft, {\sl Anderson orthogonality in the dynamics after a local quantum quench}, Phys. Rev. B {\bf 85}, 235104 (2012).


\bibitem{Ste11} J.-M. St\'{e}phan and J. Dubail, {\sl Local quantum quenches in critical one-dimensional systems: entanglement, the Loschmidt echo, and light-cone effects}, J. Stat. Mech.  P08019 (2011).

\bibitem{LuttingerParameter} A. Luther and I. Peschel, {\sl Calculation of critical exponents in two dimensions from quantum field theory in one dimension}, Phys. Rev. B {\bf 12}, 3908--3917 (1975).

\bibitem{Giamarchi} T. Giamarchi, {\sl Quantum physics in one dimension}, Oxford university press (2003).

\bibitem{AffleckGiuliano}  I. Affleck and D. Giuliano, {\sl Topological superconductor--Luttinger liquid junctions}, J. Stat. Mech, P06011 (2013).

\bibitem{LutchynSkrabacz} R.M. Lutchyn and J.H. Skrabacz, {\sl Transport properties of topological superconductor--Luttinger liquid junctions: A real-time Keldysh approach}, Phys. Rev. B {\bf 88}, 024511 (2013).

\bibitem{BCC_ND1} C.G. Callan, I.R. Klebanov, A.W.W. Ludwig, J.M. Maldacena, {\sl Exact Solution of a Boundary Conformal Field Theory}, Nucl. Phys. B {\bf 422}, 417--448  (1994).

\bibitem{BCC_ND2} P. Fendley, H. Saleur and N.P. Warner, {\sl Exact solution of a massless scalar field with a relevant boundary interaction},  Nucl. Phys. B {\bf 430},  577--596 (1994). 

\bibitem{MajoDMRG} R. Thomale, S. Rachel and P. Schmitteckert, {\sl Tunneling spectra simulation of interacting Majorana wires}, Phys. Rev. B {\bf 88}, 161103(R) (2013).

\bibitem{QuenchKondo0} R.W. Helmes, M. Sindel, L. Borda and J. von Delft, {\sl Absorption and emission in quantum dots: Fermi surface effects of Anderson excitons}, Phys. Rev. B {\bf 72}, 125301 (2005).

\bibitem{QuenchKondo1} H. E. T\"ureci, M. Han, M. Claassen, A. Weichselbaum, T. Hecht, B. Braunecker, A. Govorov, L. Glazman, A. Imamoglu and J. von Delft, {\sl Many-Body Dynamics of Exciton Creation in a Quantum Dot by Optical Absorption: A Quantum Quench towards Kondo Correlations}, 
Phys. Rev. Lett. {\bf 106}, 107402 (2011).

\bibitem{QuenchKondo3} M. Heyl and S. Kehrein, {\sl  X-ray edge singularity in optical spectra of quantum dots}, 
Phys. Rev. B {\bf 85}, 155413 (2012).

\bibitem{VM_QSH} R. Vasseur and J.E. Moore, {\sl Edge Physics of the Quantum Spin Hall Insulator from a Quantum Dot Excited by Optical Absorption}, Phys. Rev. Lett. {\bf 112}, 146804 (2014).

\bibitem{DemlerOC} M. Knap, A. Shashi, Y. Nishida, A. Imambekov, D.A. Abanin and E. Demler, {\sl Time-Dependent Impurity in Ultracold Fermions: Orthogonality Catastrophe and Beyond}, Phys. Rev. X {\bf 2},  041020 (2002).

\bibitem{GlobalLL}  B. D\'ora, F. Pollmann, J. Fort\'agh and G. Zar\'and, {\sl Loschmidt Echo and the Many-Body Orthogonality Catastrophe in a Qubit-Coupled Luttinger Liquid}, Phys. Rev. Lett. {\bf 111}, 046402 (2013).

\bibitem{QSH} M. K\"{o}nig {\sl et al.}, {\sl Quantum Spin Hall Insulator State in HgTe Quantum Wells}, Science \textbf{318}, 766 (2007).

\bibitem{Kane} C. L. Kane and E. J. Mele, {\sl Quantum Spin Hall Effect in Graphene}, Phys. Rev. Lett \textbf{95}, 226801 (2005).

\bibitem{Bernevig} B. A. Bernevig {\sl et al.}, {\sl Quantum Spin Hall Effect and Topological Phase Transition in HgTe Quantum Wells}, Science \textbf{314}, 1757 (2006).

\bibitem{Fu} L. Fu and C. L. Kane, {\sl Superconducting Proximity Effect and Majorana Fermions at the Surface of a Topological Insulator}, Phys. Rev. Lett. {\bf 100}, 096407 (2008).

\bibitem{InterplayKondoMajo} M. Cheng, M. Becker, B. Bauer, R. M. Lutchyn
, {\sl Interplay between Kondo and Majorana interactions in quantum dots}, {\tt arXiv:1308.4156}.

\bibitem{Kondo}  A. Hewson, {\sl The Kondo Problem to Heavy Fermions, Cambridge Studies in Magnetism} (Cambridge University Press, Cambridge, England, 1997).

\bibitem{DisorderedJan} D. I. Pikulin, J. P. Dahlhaus, M. Wimmer, H. Schomerus, and C. W. J. Beenakker, {\sl A zero-voltage conductance peak from weak antilocalization in a Majorana nanowire}, New J. Phys. {\bf 14}, 125011 (2012).

\bibitem{DisorderedAltland} D. Bagrets and A. Altland, {\sl Class D Spectral Peak in Majorana Quantum Wires}, Phys. Rev. Lett. {\bf 109}, 227005 (2012).

\bibitem{DisorderedDrew} J. Liu, A. C. Potter, K. T. Law, and P. A. Lee, {\sl Zero-Bias Peaks in the Tunneling Conductance of Spin-Orbit-Coupled Superconducting Wires with and without Majorana End-States}, Phys. Rev. Lett. {\bf 109}, 267002 (2012).


\bibitem{Haldane} F. Haldane, {\sl Continuum dynamics of the 1-D Heisenberg antiferromagnet: Identification with the O(3) nonlinear sigma model}, Physics Letters A {\bf 93}, 0 (1983).

\bibitem{AKLT} I. Affleck, T. Kennedy, E. H. Lieb, and H. Tasaki, {\sl Rigorous results on valence-bond ground states in antiferromagnets}, Phys. Rev. Lett. {\bf 59}, 799 (1987).

\bibitem{ReadGreen} N. Read and D. Green, {\sl Paired States of Fermions in Two Dimensions with Breaking of Parity and Time-Reversal Symmetries and the Fractional Quantum Hall Effect}, Phys. Rev. B {\bf 61}, 10 267 (2000).


\bibitem{MooreRead}  G. Moore and N. Read, {\sl Non-Abelions in the Fractional Quantum Hall Effect}, Nucl. Phys. B {\bf 360}, 362 (1991).

\bibitem{Annals} P. Fendley, M.P.A. Fisher and C. Nayak, {\sl Topological Boundary Conformal Field Theory and Tunneling of Edge Quasiparticles in non-Abelian Topological States},  Annals Phys. {\bf 324}, 1547 (2009).

\bibitem{Fendley} P. Fendley, {\sl Parafermionic edge zero modes in $\mathbb{Z}_n$-invariant spin chains}, J. Stat. Mech. (2012) P11020.

\bibitem{Para} D. J. Clarke, J. Alicea, and K. Shtengel, {\sl Exotic non-Abelian anyons from conventional fractional quantum Hall states}, Nat. Commun. {\bf 4}, 1348 (2013).

\bibitem{PRX} R.S.K. Mong, D.J. Clarke, J. Alicea, N.H. Lindner, P. Fendley, C. Nayak, Y. Oreg, A. Stern, E. Berg, K. Shtengel, and M.P.A. Fisher,  {\sl Universal topological quantum computation from a superconductor/Abelian quantum Hall heterostructure}, Phys. Rev. X {\bf 4}, 011036 (2014).


\bibitem{ReadRezayi} N. Read and E. Rezayi, {\sl Beyond Paired Quantum Hall
States: Parafermions and Incompressible States in the First Excited Landau Level}, Phys. Rev. B {\bf 59}, 8084 (1999).

\bibitem{Potts3} I. Affleck, M. Oshikawa and H. Saleur, {\sl Boundary critical phenomena in the three-state Potts model}, J. Phys. A: Math. Gen. {\bf 31}, 5827 (1998).


\end{thebibliography}
    \end{document}